\documentclass[12pt,a4paper]{article}
\usepackage{amssymb,amsmath,amsfonts,mathrsfs}
\usepackage{verbatim,amsthm,curves,graphics}
\usepackage{mathrsfs}
\usepackage{fullpage}
\usepackage{graphicx}
\usepackage{caption}
\def\ben{\begin{equation}}
\def\een{\end{equation}}
\def\bena{\begin{eqnarray}}
\def\eena{\end{eqnarray}}

\newcommand{\non}{\nonumber}

\theoremstyle{definition}
\newtheorem{thm}{Theorem}

\newtheorem{lemma}{Lemma}[section]
\newtheorem{prop}{Proposition}[section]
\newtheorem{cor}{Corollary}[section]
\newtheorem{defn}{Definition}[section]

\renewcommand{\textwidth} {16cm}

\renewcommand{\oddsidemargin} {1.5cm}

\newcommand{\de}{\delta}
\newcommand{\De}{\Delta}

\newcommand{\ka}{\kappa}
\newcommand{\la}{\lambda}

\newcommand{\vp}{\varphi}
\newcommand{\La}{\Lambda}

\newcommand{\Lao}{\Lambda_0}

\newcommand{\T}{\mathbb{T}}

\newcommand{\pa}{\partial}

\newcommand{\ti}[1]{\tilde{#1}}

\newcommand{\eq}{\begin{equation}}
\newcommand{\eqe}{\end{equation}}

\newcounter{saveeqn}

\newcommand{\mr}{{\mathbb R}}
\newcommand{\mn}{{\mathbb N}}

\newcommand{\C}{{\mathcal C}}
\newcommand{\e}{\operatorname{e}}
\newcommand{\supp}{\operatorname{supp}}

\renewcommand{\L}{{\mathcal L}}
\newcommand{\F}{{\mathscr F}}

\newcommand{\D}{{\mathcal D}}

\renewcommand{\S}{\mathscr{S}}

\renewcommand{\T}{{\mathbb T}}

\renewcommand{\O}{{\mathcal O}}

\renewcommand{\d}{{\rm d}}

\renewcommand{\D}{{\mathcal D}}

\begin{document}

\title{The operator product
expansion converges in perturbative field theory}
\author{Stefan Hollands\footnote{\ hollandss@cf.ac.uk} \\
School of Mathematics, Cardiff University, UK
\and
Christoph Kopper\footnote{\ kopper@cpht.polytechnique.fr} \\
Centre de Physique Th{\'e}orique, CNRS, UMR 7644\\
Ecole Polytechnique,
F-91128 Palaiseau, France}

\date{27 May 2011}

\maketitle

\begin{abstract}
We show, within the framework of the massive Euclidean $\varphi^4$-quantum
field theory in four dimensions,
that the Wilson operator product expansion (OPE) is not only an
asymptotic expansion at
short distances as previously believed, but even {\em converges at
  arbitrary finite} distances.
Our proof rests on a detailed estimation of the remainder term in the
OPE, of an arbitrary
product of composite fields, inserted as usual into a correlation function
with
further ``spectator fields''. The estimates are obtained
using a suitably adapted version of the method of renormalization
group flow equations.
 Convergence follows because the remainder is seen to become
 arbitrarily
small as the OPE is carried out to sufficiently high order, i.e.
to operators of sufficiently high dimension. Our results hold for
arbitrary,
but finite, loop orders. As an interesting side-result of our
estimates,
we can also prove that the ``gradient expansion'' of the effective
action is convergent.
\end{abstract}

\section{Introduction}

All quantum field theories with well-behaved ultra violet behavior
are believed to have an operator product expansion
(OPE)~\cite{Wi,Zi}.
This means that
the product of any two local fields located at nearby
points $x$ and $y$
can be expanded in the form
\ben
\label{ope1}
\O_A(x)\O_B(y) \sim \sum_C \C_{AB}^C(x-y) \, \O_C(y),
\een
where $A,B,C$ are labels for the various local fields in the
given theory (incorporating also their tensor character/spin),
and where $\C_{AB}^C$ are certain numerical coefficient
functions---or rather distributions---that depend on the theory
under consideration, the coupling
constants, etc. The
sign ``$\sim$'' indicates that this can be understood as an asymptotic
expansion: If the sum on the right side is
carried out to a sufficiently large but finite order, then the remainder
goes to zero fast as $x \to y$ in the sense of operator insertions
into a quantum state, or into a correlation function. The purpose of
this paper is to demonstrate in a specific
model that the expansion is not only {\em asymptotic} in this sense,
but even {\em converges at finite (!) distances}, to
arbitrary loop orders, in a perturbative Euclidean quantum field theory.

Our result is not merely a technical footnote, but it furnishes an
important insight into the
general structure of quantum field theory. Although our result is formulated
in a Euclidean setting, this is maybe best explained
in the Minkowskian context. There, the analogue of our result would be that
correlation functions such as the two-point function $\langle \O_A(x)
\O_B(y) \rangle_\Psi$
in a state\footnote{The state should have a well-behaved
high energy behavior. In the Minkowskian context, it should e.g. have
bounded energy $E$, see below for an
appropriate replacement in the Euclidean context.} $\Psi$ are entirely
determined by the collection of OPE
coefficients which are {\em state independent}, together with the
1-point
functions $\langle \O_C(y) \rangle_\Psi$:
\ben
\langle \O_A(x) \O_B(y) \rangle_\Psi = \sum_C \C_{AB}^C(x-y) \
\langle \O_C(y) \rangle_\Psi \, ,
\een
where the infinite sum over ``$C$'' would be convergent, and $(x-y)^2$
would not necessarily have to be
small\footnote{Note however that one expects convergence
to hold in the relativistic context only for spacelike distances,
$(x-y)^2>0$, because of locality.}.
An analogous statement would apply to the higher $n$-point functions.
Thus, the OPE coefficients capture the state-independent algebraic structure of
QFT, while {\em all} the information about the quantum state,
i.e. $n$-point functions, is
contained in the 1-point functions (``form factors'') only.
Our result is relevant
also in that it supports recent approaches to QFT such
as~\cite{Ho,HoOl,HoWa} wherein the
OPE is taken as the fundamental input.

In this paper, we prove convergence of the OPE in the context of
perturbative Euclidean QFT, to arbitrary  loop orders.
The model that we consider is
a hermitian scalar field with self-interaction $g \varphi^4$ and
mass
$m > 0$ on flat 4-dimensional Euclidean space.
The composite fields $\O_A$ in this model are simply linear combinations
of monomials in the basic field $\varphi$ and its derivatives and are
denoted by
\ben\label{compop}
\O_A = \partial^{w_1} \varphi \cdots \partial^{w_n} \varphi \, ,
\qquad A = \{n,w\} \, ,
\een
where each $w_i$ is a 4-dimensional multi-index, see ``notations and
conventions'' for
more on multi-index notation. We define the engineering dimension of such
a field as usual by
\ben
[A] = n + \sum_i |w_i|  \, .
\een
Each OPE coefficient $\C^C_{AB}(x-y)$ is itself a formal
power series in $\hbar$
(``loop expansion'').
As usual in perturbation theory, we
will not be concerned with the
convergence of these
expansions in $\hbar$. Instead, in this paper, we will be concerned
with the convergence of the OPE (i.e. the expansion in ``$C$'') at
arbitrary but fixed
order $l$ in $\hbar$.

To analyze this issue, we must insert the left- and right
sides of \eqref{ope1} into
a correlation function
containing suitable ``spectator fields'', which play the role of a
quantum state in the Euclidean context.
A simple and natural choice for the spectator fields is e.g.
\ben
\varphi(f_{p_i}) := \int d^4x\ \varphi(x) \ f_{p_i} (x) \ ,
\een
where $p_i$ is a 4-momentum, and where $f_{p_i}$ is a smooth function
whose Fourier transform $\hat f_{p_i}(q)$ has compact support for $q$ in a
ball of radius $\epsilon$ around $p_i$.
Our main result is the following

\noindent
{\bf Theorem:}
Let the sum $\sum_C$ in the operator product expansion~\eqref{ope1} be
over all $C$ such that
\ben
[C] - [A] - [B] \le \Delta
\een
where $\Delta$ is some positive integer. Then for each such $\Delta$,
we have the following bound
for the ``remainder'' in the OPE \emph{in loop order $l$:}
\bena
&& \Bigg| \bigg\langle
\O_A(x) \O_B(0) \, \varphi(f_{p_1}) \cdots \varphi(f_{p_n}) \bigg\rangle
- \sum_{C} \C_{AB}^C(x) \,
\bigg\langle \O_C(0) \, \varphi(f_{p_1}) \cdots \varphi(f_{p_n})
\bigg\rangle \Bigg|\\
&& \hspace{1.5cm} \ \ \le \ \ m^{[A]+[B]+n}\
\sqrt{[A]![B]!} \ \tilde K^{[A]+[B]}
\ \prod_i \sup |\hat f_{p_i}|
 \non\\
&& \hspace{1cm} \times  \ \
\sup(1,\frac{|\vec p|_n}{m})^{2([A]+[B])(n+2l+1)+3n}
\sum_{\lambda=0}^{n/2+2l}
\frac{\log^\lambda \sup(1,\frac{|\vec p|_n}{m})}
{2^\lambda \lambda!} \non\\
&& \hspace{2cm}
\times \ \ \frac{1}{\sqrt{\Delta!}} \ \Bigg( \tilde K \ m \ |x| \
\sup (1, \frac{|\vec p|_n}{m})^{n+2l+1} \Bigg)^{\Delta}\non \ .
\eena
Here, $\langle \, . \, \rangle$ denote
correlation functions, and $\tilde K$ is a constant depending on
$n,l$.
Furthermore,
$|\vec p|_n$ is defined in eq.~\eqref{pdef}, and $f_{p_i}$ are smooth
test functions in position space, whose support in momentum space
is contained in a ball of radius $\epsilon$ around $p_i$.

This result establishes the convergence of the OPE, i.e. the sum over $C$,
at each fixed order in perturbation theory,
because the remainder evidently goes to zero as $\Delta \to \infty$.
There are no conditions on $x$, so the OPE converges even at arbitrarily
large distances!
But we note that such conditions could arise if we were to allow
a wider class of spectator fields, for example, if we were to replace
$f_{p_i}$ by test-functions whose Fourier transforms are only decaying
in momentum space, but are not of compact support.
 This type of behavior can be understood in a way by the fact that
$|\vec p|_n$ gives a measure for
the ``typical energy'' of the ``state'' in which we try to carry out
the
OPE. As the
high energy behavior of the ``state'' becomes worse, so do the
convergence
properties
of the OPE.

\medskip
\noindent
To prove the theorem, one first has to give a prescription for
defining
the Schwinger functions and
OPE coefficients in renormalized perturbation theory. There are
several
options; in this paper we
find it convenient to use the Wilson-Wegner-Polchinski flow equation
method \cite{Po, WH, Wi}. In this method, one first
introduces an infrared cutoff called $\Lambda$, and an ultraviolet
cutoff called $\Lambda_0$. One then defines the quantities of interest
for finite values of the cutoffs, and derives for them a flow
equation as a function of $\Lambda$.
For suitable  boundary conditions its solutions may be bounded
inductively and uniformly  in the ultraviolet cutoff $\Lambda_0$.
The last fact makes it
possible to remove the cutoff\footnote{To show not only boundedness
but also convergence in the limit $\Lao \to \infty$,
one also has to study a version of the flow equation that is
differentiated w.r.t. the cutoff.
We do not perform this step here  since it has already been performed
in the literature for all quantities of interest in~\cite{KK1, KK2}.
The bounds obtained there were less precise than those
obtained here but this does not matter because also such
less stringent bounds are sufficient to merely show convergence in 
$\Lao$.} and
at the same time provides non-trivial bounds. In our case, we need
bounds for the remainder in the
OPE. Again, such bounds are verified inductively.

While the general strategy is rather clear
conceptually, it gets more involved in practice. This is because a relatively
refined induction hypothesis is required to ensure that it replicates itself
in the  induction process.
The verification of the induction step  is thus
the main technical task of this paper.

\medskip

A side result of our estimations which may be of some interest is that
the ``gradient expansion''~\eqref{gradient} of the effective action
converges at each fixed number of loops; the precise statement may be
found in Cor.~\ref{cor00}.

\paragraph{Notations and conventions:} Our convention for the Fourier
transform in $\mr^4$ is
\ben
f(x) = \int_p  \hat f(p) \e^{ipx} := \int_{\mr^4} \frac{d^4 p}{(2\pi)^4}
\e^{ipx} \hat f(p) \ .
\een
 We also use a
standard
multi-index notation.
Our multi-indices are elements $w = (w_1, \dots, w_n) \in \mn^{4n}$,
so
that each  $w_i \in \mn^4$
is a four-dimensional multiindex whose entries are $w_{i,\mu} \in \mn$
and $\mu=1,\dots,4$. If $f(\vec p)$ is a smooth function on $\mr^{4n}$, we set
\ben
\pa^{w} f(\vec p) = \prod_{i,\mu}
\left( {\pa \over \pa p_{i,\mu}} \right)^{w_{i,\mu}} f(\vec p)
\een
and
\ben
w! = \prod_{i,\mu} w_{i,\mu}! \, , \quad |w|=\sum_{i,\mu} w_{i,\mu} \, .
\een
We often need to take derivatives $\partial^w$ of a product of
functions
$f_1 \dots f_n$.
Using the Leibniz rule, such derivatives get distributed over the
factors
resulting
in the sum of all terms of the form $c_{\{v_i\}} \ \partial^{v_1} f_1
\dots \partial^{v_r} f_r$, where
each $v_i$ is now a $4n$-dimensional multi-index, where
$v_1+\dots+v_r=w$,
and where
\ben
c_{\{v_i\}} = \frac{(v_1+\dots+v_r)!}{v_1! \dots v_r!} \le r^{|w|}
\een
is the associated weight factor.

If $F(\varphi)$ is a differentiable function (in the Frechet space
sense)
of the
Schwartz space function $\varphi \in \S(\mr^4)$, we denote its
functional
derivative as
\ben
\frac{\d}{\d t} F(\varphi + t\psi) |_{t=0} = \int d^4 x \
\frac{\delta F(\varphi)}{\delta \varphi(x)} \ \psi(x) \ ,
\quad \psi \in \S(\mr^4)\ ,
\een
where the right side is understood in the sense of distributions in
$\S'(\mr^4)$. Multiple functional derivatives are
denoted in a similar way and define in general distributions on
multiple cartesian copies of $\mr^4$.

\section{Basic setup, flow equation framework}
\label{sec2}

In this section we introduce the quantities of interest in this paper,
namely the Schwinger functions $\langle \dots \rangle$, and the OPE
coefficient functions,
$\C^C_{AB}$. For this purpose, we will also derive various useful
auxiliary quantities such as
connected and amputated Schwinger functions, as well as certain
``normal products''. The reason
for defining these is that they satisfy a suitably simple version
of the flow equations, which we also give below.
Renormalization theory based on the  flow equation (FE) \cite{WH, Wi, Po}
of the renormalization group
has been reviewed quite often in
the literature, so we will be relatively brief.
The first presentation in the form we use it here is in \cite{KKS}.
Reviews are in \cite{Mu} and in  \cite{Kop} (in German).

\subsection{Connected amputated Green functions (CAG's)}

To begin, we introduce an infrared\footnote{Such a cutoff is of
course not necessary
in a massive theory. The IR behavior is substantially modified only
for $\La$ above
$m$.} cutoff $\Lambda$, and an ultraviolet cutoff $\Lambda_0$. These
cutoffs enter the
definition of the theory through the propagator
$C^{\Lambda,\Lambda_0}$
which is defined in
momentum space by
\eq
C^{\La,\Lao}(p)\,=\, {1 \over  p^2+m^2}
\left[ \exp \left(- {p^2+m^2 \over \Lao^2} \right) - \exp
\left(- {p^2+m^2 \over \La^2} \right) \right] \, .
\label{propreg}
\eqe
The full propagator is recovered for $\La \to 0$ and $\Lao \to \infty\,$,
and we always assume
\eq
\label{ka}
0<\Lambda\ , \quad \ka := \sup(\La,m) < \Lao \ .
\eqe
Other choices of regularization are of course admissible. The one
chosen in (\ref{propreg}) has the advantage of being analytic in $p^2$
for $\La>0$.
The propagator defines a corresponding
Gaussian measure
$\mu^{\La,\Lao}$, whose covariance is $\hbar C^{\La,\Lao}$.
The factor of $\hbar$ is inserted to obtain a consistent
loop expansion in the following.
The interaction is taken to be
\ben
L^{\Lambda_0}(\varphi) = \int d^4 x \ \bigg( a^{\Lambda_0}
\, \varphi(x)^2
+b^{\Lambda_0} \, \partial \varphi(x)^2+c^{\Lambda_0}
\, \varphi(x)^4 \bigg) \ .
\label{ac}
\een
It contains suitable counter terms satisfying
$a^{\Lambda_0} = O(\hbar),\  b^{\Lambda_0} = O(\hbar^2)$
and $c^{\Lambda_0} = \frac{g}{4!} +
O(\hbar)$. They will be
adjusted--and actually diverge--when $\Lambda_0 \to \infty$
in order to obtain a well
defined limit of the quantities of interest for us. We have anticipated this
by making them ``running couplings'', i.e. functions of the ultra
violet cutoff $\Lambda_0$.
The correlation ($=$ Schwinger-) functions of $n$ basic fields with
cutoff are then given by
\ben\label{pathint}
\langle \varphi(x_1) \cdots \varphi(x_n) \rangle :=
(Z^{\Lambda,\Lambda_0})^{-1}
\int d\mu^{\Lambda,\Lambda_0} \ \exp \bigg( -\frac{1}{\hbar}
L^{\Lambda_0}\bigg) \, \varphi(x_1) \cdots \varphi(x_n) \, .
\een
This is just the standard Euclidean path-integral, but note that
the free part in the Lagrangian
has been absorbed into the Gaussian measure $d\mu^{\La,\Lao}$.
The normalization factor
is chosen so that $\langle 1 \rangle = 1$. This factor is finite only
as long as we impose an additional volume cutoff. But the infinite volume limit
can be taken without difficulty once we pass to perturbative
connected correlation functions which we will do in a moment.
For more details on this limit see~\cite{KMR, Mu}.
 The path integral will be analyzed in
the perturbative sense, i.e. the exponentials are expanded out and the
Gaussian integrals are then performed.
The full theory is obtained by sending the cutoffs $\Lambda_0 \to
\infty$
and $\Lambda \to 0$,
for a suitable choice of the running couplings.
In the flow equation technique, the
correct behavior of the running
couplings, necessary for a well-defined limit, is obtained by deriving
first a differential equation for the Schwinger functions in
$\Lambda$,
and by then defining the running couplings implicitly through the boundary
conditions for this equation.

These flow equations are written more conveniently in
terms of the hierarchy of ``connected, amputated
Schwinger functions'' (CAG's).
Their generating functional is defined  through the
convolution\footnote{The convolution is defined in general by
$(\mu^{\Lambda,\Lambda_0} \star F)(\varphi) =
\int d\mu^{\Lambda,\Lambda_0}(\varphi') \ F(\varphi+\varphi')$.}
of the Gaussian measure with the exponentiated interaction.
\ben\label{CAGdef}
-L^{\Lambda, \Lambda_0} := \hbar \, \log \, \mu^{\Lambda,\Lambda_0}
\star \exp \bigg(-\frac{1}{\hbar} L^{\Lambda_0}
\bigg)- \hbar \log  Z^{\La,\Lao} \ .
\een
The functional
$L^{\Lambda,\Lambda_0}$ has an expansion
as a formal power series
in terms of Feynman diagrams with precisely $l$ loops, $n$ external
legs, and propagator $C^{\Lambda,\Lambda_0}(p)$. As the name suggests,
only connected diagrams contribute,
and the (free) propagators on the external legs are removed. We will
not use  decompositions in terms of Feynman diagrams.
But we will also analyze the functional (\ref{CAGdef})
in the sense of formal power
series, i.e. we consider the terms in the formal power series
\ben\label{genfunc}
L^{\Lambda, \Lambda_0}(\varphi) := \sum_{n>0}^\infty
\sum_{l=0}^\infty {\hbar^l}
\int d^4x_1 \dots d^4 x_n\ \L^{\Lambda,\Lambda_0}_{n,l}(x_1, \dots, x_n)
\,
\varphi(x_1) \cdots \varphi(x_n) \, ,
\een
where $\varphi \in \S(\mr^4)$ is any Schwartz space function. No
statement
is made about the
convergence of the series in $\hbar$.
The objects on the right side, the CAG's,
 are the basic quantities in our analysis
because they are
easier to work with than the full Schwinger functions. But the
latter can of course be recovered from the CAG's.

Because the connected amputated functions in position space are translation
invariant, their Fourier transforms, denoted
$\L^{\La,\Lao}_{n,l}(p_1, \dots, p_n)$, are supported
at $p_1+\dots+p_n=0$.
We consequently write, by
abuse of notation
\ben
\L^{\Lambda,\Lambda_0}_{n,l}(p_1, \dots, p_n) = \delta^{4}{(\sum_{i=1}^n
p_i)}
\, \L^{\Lambda,\Lambda_0}_{n,l}(p_1, \dots, p_{n-1}) \, ,
\een
i.e. one of the momenta is determined in terms of the remaining $n-1$
independent momenta by momentum conservation. It is straightforward to
see that, as
functions of these remaining independent momenta, the connected
amputated
Green functions
are smooth, $\L^{\Lambda,\Lambda_0}_{n,l}(p_1, \dots, p_{n-1})
\in C^\infty(\mr^{4(n-1)})$. It is much
less obvious, but will be demonstrated later in Cor.~\ref{cor00},
that they are in fact
even analytic functions near $\vec p=0$,
even after the cutoffs are removed.

The flow equations are
obtained by taking a $\La$-derivative of
eq.\eqref{CAGdef}:
\ben
\partial_{\La} L^{\La,\Lao} \,=\,
\frac{\hbar}{2}\,
\langle\frac{\delta}{\delta \vp},\dot {C}^{\La}\star
\frac{\delta}{\delta \vp}\rangle L^{\La,\Lao}
\,-\,
\frac{1}{2}\, \langle \frac{\delta}{\delta
  \vp} L^{\La,\Lao},
\dot {C}^{\La}\star
\frac{\delta}{\delta \vp} L^{\La,\Lao}\rangle  +
\hbar \partial_\Lambda \log Z^{\La,\Lao} \ .
\label{fe}
\een
Here we use the shorthand $\,\dot {C}^{\La}\,$ for
$\partial_{\La} {C}^{\La,\Lao}\,$, which, as we note,
does not depend on $\Lao$.
By $\langle\ ,\  \rangle$ we denote the standard scalar product in
$L^2(\mathbb{R}^4, d^4 x)\,$, and $\star$ denotes
convolution in $\mr^4$. For example
\ben
\langle\frac{\delta}{\delta \vp},\dot {C}^{\La}\star
\frac{\delta}{\delta \vp}\rangle = \int d^4x d^4y \ \dot {C}^{\La}(x-y)
\frac{\delta}{\delta \varphi(x)} \frac{\delta}{\delta \varphi(y)}
\een
is the ``functional Laplace operator''.
When expanded out in $\varphi$, the flow equations~\eqref{fe}
read in momentum space
\bena
&&\pa_{\La} \, {\cal L}^{\La,\Lao}_{2n,l}(p_1,\ldots p_{2n-1}) =
\left( {2n+2 \atop 2} \right) \int_k {\dot C}^{\La}(k)\,
{\cal L}^{\La,\Lao}_{2n+2,l-1}(k,-k, p_1,\ldots p_{2n-1})
\label{feq00} \non \\
&&
-2\!\!\!\!\!\!\!
\sum_{\begin{array}{c}_{l_1+l_2=l,}\atop
_{n_1+n_2=n +1}\end{array} }\!\!\!\!\!\!\!\! n_1n_2 \,
{\mathbb S} \Biggl[ {\cal
  L}^{\La,\Lao}_{2n_1,l_1}(p_1,\ldots,p_{2n_1-1})\,
{\dot C}^{\La}(q)\,\,
{\cal L}^{\La,\Lao}_{2n_2,l_2}(-q,p_{2n_1},\ldots,p_{2n-1})
\Biggr]
\label{feqq}
\eena
\[ \mbox{with }\quad
q= -p_1 -\ldots -p_{2n_1-1}\,= \,p_{2n_1}
+ \,p_{2n_1+1} +\ldots +p_{2n}\ .
\]
The symbol ${\mathbb S}$ is an operator which acts on the functions of
momenta $(p_1, \dots, p_{2n})$ by
taking the mean value over those permutations $\pi\,$
of $(1,\ldots, 2n)\,$, for which
$\pi(1) < \pi(2) <\ldots < \pi(2n_1-1)\,$ and
$\,\pi(2n_1) < \pi(2n_1+1) < \ldots < \pi(2n)\,$.
And we used the fact that for the theory proposed through
(\ref{ac}), only even moments of the effective action
will be non-vanishing due to the symmetry $\vp \to -\vp\,$,
and we thus wrote the equations only for those.\\
We will also need the FE's differentiated w.r.t. to
components of the momentum variables. We
obtain\footnote{In distributing the derivatives over
the three
  factors in the second term on the r.h.s. with the Leibniz rule,
we have tacitly assumed
  that the momentum $p_i$ appears among those from
${\cal  L}^{\La,\Lao}_{2n_1,l_1}\,$. If this is not the case
one has to parametrize  ${\cal L}^{\La,\Lao}_{2n_1,l_1}\,$
in terms of (say) $(p_2,\ldots p_{2n_1-1}, q)\,$ with
$\,q\,= \,p_{2n_1}+\ldots +p_{2n}\,$, in order to introduce the
 $p_i$-dependence in   $ {\cal L}^{\La,\Lao}_{2n_1,l_1}\,$.
For a  fully systematic treatment see \cite{GK}.}:
\bena
&&\pa_{\La} \pa^{w} \, {\cal L}^{\La,\Lao}_{2n,l}(p_1,\ldots, p_{2n-1}) =
\left( {2n+2 \atop 2} \right) \int_k {\dot C}^{\La}(k)\,\pa ^w
{\cal L}^{\La,\Lao}_{2n+2,l-1}(k,-k, p_1,\ldots p_{2n-1})
\label{feq0} \\
&&-2\!\!\!\!\!\!\!\!\!\!\!\!\sum_{
{\tiny
\begin{array}{c}
l_1+l_2=l\\
w_1+w_2+w_3=w\\
n_1+n_2=n +1
\end{array}
}
}
\!\!\!\!\!\!\!\!\!\!\!\!
n_1 n_2 \ c_{\{w_j\}} \, \mathbb{S} \Biggl[ \pa^{w_1} {\cal
  L}^{\La,\Lao}_{2n_1,l_1}(p_1,\ldots,p_{2n_1-1})
\,\pa^{w_3} {\dot C}^{\La}(q)\,
\pa^{w_2}{\cal L}^{\La,\Lao}_{2n_2,l_2}(-q, p_{2n_1},\ldots,p_{2n-1})
\Biggr] \ .\non
\eena

To define the  CAG's through the flow equations,
we have to impose  boundary
conditions. These are\footnote{We restrict to BPHZ renormalization
conditions in their simplest form, more general choices are of
course equally admissible.}, using
the multi-index convention introduced above in ``notations and conventions'':
\ben
\partial^w \L^{0,\Lambda_0}_{n,l}(\vec 0) = \de_{w,0}\ \de_{n,4}\
 \de_{l,0}\ \frac{g}{4!} \quad \text{for $n+|w|
\le 4$,}
\een
 as well as
\ben
\partial^w \L^{\Lambda_0,\Lambda_0}_{n,l}(\vec p) = 0 \quad \text{for
$n+|w| > 4$.}
\een
The  CAG's are then determined by integrating the
flow equations subject to these boundary conditions, see~\cite{KKS,Mu}.
In our context this is described in detail
when we come to the estimates of the CAG's in sec.~\ref{sec3}.

\subsection{Insertions of composite fields,
normal products and OPE coefficients}

For the purposes of this paper, and also in many applications,
one would like to define not only
Schwinger functions of products of the basic field, but also ones
containing composite operators.
These are obtained by replacing the action $L^{\Lambda_0}$ with an
action containing additional sources.
To set things up properly, it is useful to introduce first some
notation.
Let $\F_{{\rm loc}}^\infty$ be the space of
smooth local, polynomial functionals $F(\varphi)$ of
$\varphi \in \S(\mr^4)$. Any
such functional can be
written by definition as
\ben
F (\varphi)= \sum_A \int d^4 x \  \O_A(x) \ f^A(x) \,\, ,
\quad f^A \in C^\infty_0(\mr^4) \, ,
\een
where $\O_A$ are composite operators as in eq.~\eqref{compop} and
where
the sum is finite.
We now consider instead of $L^{\Lambda_0}$ a modified action
containing
sources $f^A$, given by
replacing
\ben
L^{\Lambda_0}\to L^{\Lao}_F:=L^{\Lambda_0}+ F + \sum_{j=0}^\infty
B^{\Lambda_0}_j(\underbrace{F \otimes \cdots \otimes F}_j) \, ,
\een
where the last term represents the counter terms and is for each $j$ a
suitable linear functional\footnote{As we will see it is possible to
 impose boundary conditions such that the multiply inserted
Schwinger functions become less singular at short distances 
$(x_i-x_j)^2 \to 0$.
In this case the maps $B_j$ take their values in the space 
$\mathscr{F}^\infty$ of {\em non-local} 
 functionals on Schwartz space.}
\ben
B_j^{\Lambda_0}: (\F_{{\rm loc}}^\infty)^{\otimes j}
\to \F^\infty
 \  ,
\een
that is symmetric, and of order $O(\hbar)$.
These counter terms
are designed to eliminate the divergences arising from composite field
insertions in the
Schwinger functions when one takes $\Lambda_0 \to \infty$.
The Schwinger functions with insertions of $r$~composite operators are
defined with the aid of  functional
derivatives with respect to the sources, setting the sources
$f^{A_i} =0$ afterwards:
\ben
\langle \O_{A_1}(x_1) \cdots \O_{A_r}(x_r)  \rangle :=
\een
\[
\hbar^r
\frac{\delta^r}{\delta f^{A_1}(x_1) \dots \delta f^{A_r}(x_r)}
\ (Z^{\La,\Lao})^{-1} \int d\mu^{\Lambda,\Lambda_0}
\exp \bigg(-\frac{1}{\hbar}  L^{\Lambda_0}_F(\varphi)
\bigg)\biggr|_{ f^{A_i}=0} \, .
\]
The previous definition of the CAG's is a special case
of this; there we
take $F = \int d^4 x \ f(x) \ \varphi(x)$, and we have
$B^{\Lambda_0}_j(F^{\otimes j})=0$, because no extra counter terms
are required for this simple insertion. As above, we can define
a corresponding effective action
as
\ben\label{genfuncins}
-L^{\Lambda,\Lambda_0}_F := \hbar \, \log \, \mu^{\Lambda,\Lambda_0}
\star \exp \bigg(-\frac{1}{\hbar} ( L^{\Lambda_0}
+ F + \sum_{j=0}^\infty B^{\Lambda_0}_j(F^{\otimes j})) \bigg)
- \log Z^{\La,\Lao}
\een
which is now a functional of the sources $f^{A_i}$, as well as of $\varphi$.
Differentiating $r$ times with respect to the sources, and setting them
to zero afterwards, gives the generating functionals of the CAG's with
$r$ operator insertions, namely:
\ben
L^{\Lambda,\Lambda_0}(\O_{A_1}(x_1) \otimes \dots \otimes \O_{A_r}(x_r))
=
\frac{\delta^r \ L^{\Lambda,\Lambda_0}_F}{\delta f^{A_1}(x_1) \dots \delta
  f^{A_r}(x_r)}
\,  \Bigg|_{f^{A_i} =  0} \, .
\een
The CAG's with insertions satisfy a number of obvious properties,
e.g.
they are multi-linear--as indicated by the
tensor product notation--and symmetric in the insertions.

As the CAG's without insertions, the CAG's with insertions can be further
expanded in $\varphi$ and $\hbar$, and this is denoted as
\ben
L^{\Lambda,\Lambda_0} \bigg( \bigotimes_{i=1}^r \O_{A_i}(x_i) \bigg) =
\sum_{n,l \ge 0} {\hbar^l} \int d^4y_1\dots d^4y_n \
\L^{\Lambda,\Lambda_0}_{n,l}\bigg( \bigotimes_{i=1}^r \O_{A_i}(x_i);
y_1,
\dots, y_n \bigg)
\prod_{j=1}^n \vp(y_j) \, .\non
\een
Due to the insertions in $\L_{n,l}^{\La,\Lao} (\otimes_j
\O_{A_j}(x_j),
\vec p)$, there is no restriction on the momentum set
$\vec p$. However it follows from translation invariance that functions
with insertions at a translated set of points $x_j + a\,$ are
obtained
from those
at $a=0\,$ upon multiplication by \eq
\e^{ia\sum_{i=1}^n p_i}\ .
\eqe
The CAG's with insertions satisfy a flow equation of a similar
nature as those without insertions, and these are again
obtained by taking derivatives  of
eq.~\eqref{fe} with respect to the sources.
For example, for one insertion, the flow equation (FE) is:
\eq
\partial_{\La}\, L^{\La,\Lao}(\O_A) \,=\,
\frac{\hbar}{2}\,
\langle\frac{\delta}{\delta \vp},\dot {C}^{\La}\star
\frac{\delta}{\delta \vp}\rangle L^{\La,\Lao}(\O_A)
\,-\,
\langle \frac{\delta}{\delta
  \vp} L^{\La,\Lao}(\O_A) ,
\dot {C}^{\La}\star
\frac{\delta}{\delta \vp} L^{\La,\Lao}\rangle \,+\,
\partial_{\La}\, \log Z^{\La,\Lao} \, .
\label{feqin}
\eqe
It is important to note that this FE is {\it linear and homogeneous}
w.r.t. the functional
$L^{\La,\Lao}(\O_A)\,$. The FE's for multiple insertions are obtained
similarly by taking more functional derivatives with respect to the
sources, for example for two insertions:
\bena
\label{feqin1}
\partial_{\La}\, L^{\La,\Lao}(\O_A \otimes \O_B) &=&
\frac{\hbar}{2}\,
\langle\frac{\delta}{\delta \vp},\dot {C}^{\La}\star
\frac{\delta}{\delta \vp}\rangle L^{\La,\Lao}(\O_A \otimes \O_B)\non\\
&-&
\langle \frac{\delta}{\delta
  \vp} L^{\La,\Lao}(\O_A \otimes \O_B) ,
\dot {C}^{\La}\star
\frac{\delta}{\delta \vp} L^{\La,\Lao}\rangle \, \\
&-&  \langle \frac{\delta}{\delta
  \vp} L^{\La,\Lao}(\O_A) ,
\dot {C}^{\La}\star
\frac{\delta}{\delta \vp} L^{\La,\Lao} (\O_B) \rangle \,+\,
\partial_{\La}\, \log Z^{\La,\Lao} \, .\non
\eena
Thus, the flow equation for the CAG's with two insertions is not
linear homogeneous, but involves a ``source term''
which is quadratic in the CAG's with one insertion. If we want to
integrate the flow
equations with insertions, we therefore have to ascend in the number
of insertions.

Expanding the FE's for the generating functionals
in terms of $\hbar$  and $\varphi$ gives us  again a corresponding
hierarchy of FE's satisfied by the
$\L^{\La,\Lao}_{n,l}(\otimes_i \O_{A_i}; \vec p)$.
For one insertion, these equations are given below
in eq.~\eqref{fe1ins}, whereas for
two insertions, they are given below in
eq.~\eqref{fe2ins}
(without the index ``$D$'').

The  CAG's with one insertion are not uniquely defined without
imposing suitable boundary conditions on the corresponding FE.
For an operator $\O_A$ and $A = \{n', w'\}$
(so that its dimension is $[A] = n' + |w'|$)
the simplest choice of boundary conditions, which also goes under
the name of ''normal product'', is
\eq
\partial^w {\cal L}^{\Lao,\Lao}_{n,l}(\O_A(0); \vec p) = 0
\quad \text{for} \quad
n+|w| > n'+|w'| \ ,
\label{br1}
\eqe
and
\eq
\partial^w  {\cal L}^{0,\Lao}_{n,l}(\O_A(0); 0) = i^{|w|} w! \,
\delta_{w,w'} \delta_{n,n'} \delta_{l,0}
\quad \text{for} \quad
n+|w| \leq n' + |w'|\,.
\label{br2}
\eqe
The $\delta$-symbol only depends on the {\em sets}
$\{w\}=\{w_1,\ldots,w_{n}\}$ and $\{w'\}=\{w'_1,\ldots,w'_{n'}\}$.\\
Due to the linearity of the FE, linear superpositions of normal
products are also solutions of the system of FE, and their
boundary values are the corresponding superpositions.\\
It is also possible to extend the definition of the normal products
in the following sense which leads to the appearance of an additional
index, $D$,  measuring the degree of regularity.
For one insertion, these ``oversubtracted''  normal products are denoted
$\L^{\La,\Lao}_{n,l,D}(\O_A; \vec{p})$
and are defined for any $D \ge [A]$ through
\eq
\partial^w {\cal L}^{\Lao,\Lao}_{n,l,D}(\O_A(0); \vec p) = 0
\quad \text{for} \quad
n+|w| > D  \ ,
\label{br11}
\eqe
and
\eq
\partial^w  {\cal L}^{0,\Lao}_{n,l,D}(\O_A(0); \vec{0})
= m^{D-n-|w|} i^{|w|} w! \, \delta_{w,w'} \delta_{n,n'} \delta_{l,0}
\quad \text{for} \quad
n+|w| \leq D\,.
\label{br21}
\eqe
In particular, for $D=n'+|w'|=[A]$, the oversubtracted
 normal products agree with the previous ones, because they
then satisfy the same FE and the same boundary conditions.

As the CAG's with one insertion, the CAG's with multiple insertions
are not uniquely defined by the FE without imposing a boundary condition.
The simplest boundary conditions for two insertions are given by:
\eq
\partial^w {\cal L}^{\Lao,\Lao}_{n,l}(\O_A(x) \otimes \O_B(0); \vec p) = 0
\quad \text{for all} \quad
n+|w| \ge 0 \ ,
\label{br3}
\eqe
and for all $A,B$. Imposing these boundary conditions means that
no regularizing counter terms for the corresponding operator
product are introduced.
The FE's for the CAG's with insertions may be integrated
subject to these boundary conditions, and this will be our prescription for
actually defining them. In the end, the cutoffs $\La,\Lao$ are taken
away, and the limits will be controlled by the estimates that are given in
the next section~\ref{sec3}.\\
Regularized
operator products for two or more insertions are denoted
$\L^{\La,\Lao}_{n,l,D}(\otimes_i \O_{A_i}; \vec{p})$
and are defined for any $D \ge 0$. They are defined as the
solutions to the FE~\eqref{feqin1}, together with the boundary conditions
\eq
\partial^w  {\cal L}^{0,\Lao}_{n,l,D}(\otimes_i \O_{A_i}(x_i); \vec{0}) = 0
\quad \text{for} \quad
n+|w| \leq D\ ,
\label{br2a}
\eqe
\eq
\partial^w  {\cal L}^{\Lao,\Lao}_{n,l,D}(\otimes_i \O_{A_i}(x_i); \vec{p}) = 0
\quad \text{for} \quad
n+|w| > D\ .
\label{br2b}
\eqe
In particular, for $D=-1$, the normal products agree with the
previously
defined CAG's with multiple insertions, because they then satisfy the
same
boundary conditions and FE.

A useful property of the CAG's (both `standard' and `oversubtracted'), which
follows from our choice of boundary conditions, is the
following. Let $\O_A$ be as usual a
monomial
in $\varphi$ and its derivatives. Furthermore, for any multi-index
$w \in \mn^4$,
let $\partial^w \O_A$ be the linear combination of monomials that are
obtained by carrying out the derivatives
in the obvious way.
Then the CAG's are seen~\cite{KK1, KK2}
to satisfy the ``Lowenstein rule'':
\bena
&& \partial_{x_i}^w \,
L^{\Lambda,\Lambda_0}_D( \O_{A_1}(x_1) \otimes \dots \otimes \O_{A_r}(0)) \\
&=&
\begin{cases}
L^{\Lambda,\Lambda_0}_{D}(\O_{A_1}(x_1) \otimes \dots \partial_{x_i}^w
\O_{A_i}(x_i) \otimes
\dots \O_{A_r}(0) ) & \text{for $r \ge 2$, $i \le r-1$, $D \ge 0$,}\\
L^{\Lambda,\Lambda_0}_{D+|w|}(\partial_{x_1}^w\O_{A_1}(x_1)) &
\text{for $r=1$, $i=1$, $D\ge [A_1]$} \ .
\end{cases}
\non
\eena
This property is important in order to define insertions containing
derivatives in a consistent way
and has also been termed ``action Ward identity'', or ``Leibniz rule''.
See~\cite{FreDu,HoWa1} for a discussion of such conditions in other
setups of renormalization theory.

A major advantage of the CAG's for our purposes is
that the OPE coefficients
can be expressed in terms of them in relatively simple manner,
as we now explain.
For $F(\varphi)$ a differentiable functional
of Schwartz space functions $\varphi \in \S(\mr^4)$, let $\D^A$ be the
operator defined as
\ben
\D^A F 
= \frac{(-i)^{|w|}}{n! \ w!} \ \partial^w_{\vec p} \frac{\delta^n}{\delta
\hat \varphi(p_1) \cdots \delta \hat \varphi(p_n)} \ F(\varphi) \
\Bigg|_{\hat \varphi=0, \vec p=0} \, \quad ,
\quad \text{where $A=\{n,w\}$.}
\een
Also, for a sufficiently smooth function $f$ on $\mr^4$, let the Taylor
expansion operator $\T^j$ be defined as
\ben\label{taylor}
\T^j f(x) = \sum_{|w|=j} \frac{x^w}{w!} \partial^w f(0) \ .
\een
Then the OPE coefficients are defined as follows:

\begin{defn}
For a finite UV-cutoff $\Lao$, the OPE coefficients, $\C^C_{AB}(x)$ are
defined as follows:
\begin{enumerate}
\item Let $[C]-[A]-[B] < 0$. Then we define
\ben
\C^C_{AB}(x) := \D^C \ \Bigg\{ \hbar L^{0,\Lao}_{[C]-1} (\O_A(x)
\otimes \O_B(0)) \Bigg\} \ .
\label{45}
\een
\item Let $[C]-[A]-[B] \ge 0$. Then we define
\bena
\C^C_{AB}(x) &:=& \D^C \ \Bigg\{ \hbar L^{0,\Lao}_{[C]-1}
((1-\Sigma_{j=0}^{[C]-[A]-[B]-1} \T^j) \O_A(x) \otimes \O_B(0)) - \non\\
&& \ \ \ - \ L^{0,\Lao}_{[C]-[B]}
(\T^{[C]-[A]-[B]} \O_A(x)) \ L^{0,\Lao}_{[B]} (\O_B(0)) \Bigg\} \ .
\label{46}
\eena
\end{enumerate}
\end{defn}
Our bounds in the subsequent sections, or those in~\cite{KK2}, imply
that we can remove
the cutoff $\Lao$ in the CAG's in the above formulas, and that the
$\C^C_{AB}$ are
well-defined (as smooth functions for $x \in \mr^4 \setminus \{0\}$)
in the limit as $\Lao \to \infty$. The OPE coefficients of the theory
without
cutoffs are defined to be this limit.

For our analysis of the operator products we need the FE's
expanded w.r.t. the number of fields  and loops.
For one insertion we obtain from ~\eqref{feqin}:
\eq
\pa_{\La}  \, {\cal L}^{\La,\Lao}_{2n,l}(\O_A;p_1,\ldots, p_{2n}) =
\left( {2n+2 \atop 2} \right) \int_k {\dot C}^{\La}(k)\,
{\cal L}^{\La,\Lao}_{2n+2,l-1}(\O_A; k,-k, p_1,\ldots p_{2n})
\label{fe1ins}
\eqe
\[
-4\!\!\!\!\!\!\!\!
\sum_{
{\tiny
\begin{array}{c}
l_1+l_2=l\\
n_1+n_2=n +1
\end{array}
}
}
\!\!\!\!\!\!\!\! \!\!\!\! n_1 n_2 \
 \mathbb{S} \Biggl[
{\cal L}^{\La,\Lao}_{2n_1,l_1}(\O_A; q, p_1,\ldots,p_{2n_1-1})
 {\dot C}^{\La}(q)\,
{\cal  L}^{\La,\Lao}_{2n_2,l_2}(\, p_{2n_1},\ldots,p_{2n})
\Biggr]
\]
${ \qquad\qquad}$
with\footnote{Note that by symmetry and translation invariance
${\cal  L}^{\La,\Lao}_{2n_2,l_2}(\, p_{2n_1},\ldots,p_{2n})$
$={\cal  L}^{\La,\Lao}_{2n_2,l_2}(\,- q,p_{2n_1},\ldots,p_{2n-1})$.}
$q \,= \, \,p_{2n_1}
\,+ \ldots +\, p_{2n}\ .$\\
When expanded out in moments and powers of
$\hbar$ the FE's for two insertions ~\eqref{feqin1}
read:
\bena
\label{fe2ins}
&&\pa_{\La} \, {\cal L}^{\La,\Lao}_{2n,l,D}(\O_A \otimes \O_B; p_1,
\ldots, p_{2n}) \\
&&\hspace{7mm}\non\\
&&=
\left( {2n+2 \atop 2} \right) \int_k {\dot C}^{\La}(k)\,
{\cal L}^{\La,\Lao}_{2n+2,l-1,D}(\O_A \otimes \O_B; k,-k, p_1,\ldots,
p_{2n}) \non\\
&&\hspace{4mm}\non\\
&&-4\!\!\!\!\!\!\!
\sum_{
{\tiny
\begin{array}{c}
l_1+l_2=l,\\
n_1+n_2=n+1
\end{array}
} }
n_1 n_2 \
{\mathbb S} \Biggl[ {\cal
  L}^{\La,\Lao}_{2n_1,l_1,D}(\O_A \otimes
\O_B; q, p_1,\ldots,p_{2n_1-1})\,
{\dot C}^{\La}(q)\,\,
{\cal L}^{\La,\Lao}_{2n_2,l_2}(p_{2n_1},\ldots,p_{2n})
\non\\
&&\qquad
 -\int_k {\cal
  L}^{\La,\Lao}_{2n_1,l_1}(\O_A; k, p_1,\ldots,p_{2n_1-1})\,
{\dot C}^{\La}(k)\,\,
{\cal L}^{\La,\Lao}_{2n_2,l_2}(\O_B; -k, p_{2n_1},\ldots,p_{2n})
\Biggr]\non
\eena
\[ \mbox{with }\quad
q= \,p_{2n_1}
\,+ \ldots +\, p_{2n} \, .
\]
The symmetrization operator ${\mathbb S}$ is defined as above in~\eqref{feqq}.

\section{Bounds on CAG's}
\label{sec3}

In this section, we will derive bounds on the CAG's, including those with
insertions. These bounds will imply the existence of the
limits $\Lambda \to 0$ and $\Lambda_0 \to \infty$, but they will also be
sufficient  to prove the main result Thm.~\ref{remainder}
of this paper.

The bounds on the CAG's depend on the choice of the coupling
constant $g\,$ entering the flow equation via the boundary condition
$\L_{4,0}^{0,\Lambda_0}(\vec 0) = \frac{g}{4!}$. The loop expanded
(inserted or non inserted) CAG's depend on this
coupling in an obvious way;  the noninserted functions
${\cal L}_{2n,l}\,$
carry a power of  $g^{\frac{2n-2}{2}+l}\,$ for example. To simplify the
subsequent bounds, we will always set $g=1$ in the following.

\subsection{A collection of useful bounds}

The following bounds  which largely stem from~\cite{Ko}~
will be useful to control the solutions of the various FE's:\\
The $\La$-derivative of the propagator (\ref{propreg})
is given by
\eq
{\dot C}^{\La}(p) = -\frac{2}{\La ^3} \
\e^{-\frac{p^2+m^2}{\La ^2}}\ .
\label{33}
\eqe
We find\\
1)
\eq
2 \frac{{m^3}}{\La ^3} \
\e^{-\frac{m^2}{\La ^2}}\, \le \,
1\ ,\quad
 \quad
\frac{{m^N} }{\La ^N} \
\e^{-\frac{m^2}{\La ^2}}\ \le \
\sqrt{N!}\ .
\label{la4}
\eqe

\noindent
2) For given momentum set $(p_1,\ldots,p_{n})$
we use the (shorthand) definitions
\eq
\vec{p} \equiv (p_1,\ldots,p_{n})\ ,\quad
|\vec{p}|_{n} \equiv\sup_{J \subset \{1,\ldots,n\}}
 |\sum_{i\in J} p_i|\ ,\quad
\vec{p}_{n+2} \equiv
(\vec{p}, \,k,\, -k)\ .
\label{pdef}
\eqe
Subsequently we sometimes simply write $|\vec p|$ instead of $|\vec p|_{2n}$.
Then we claim
\eq
\int_{\frac{k}{\La}}  \e^{-\frac12 (\frac{k}{\La})^2}\
\log^{\la}(\sup({|\vec{p}|_{2n+2} \over \ka},{\ka\over m}))\ \le \
\log^{\la}(\sup({|\vec{p}| \over \ka},{\ka\over m}))
\,+\,   [\la!]^{1/2}\ , \quad
\ka = \sup(\La,m)\ .
\label{momint}
\eqe
The proof is in~\cite{Ko}~, Lemma~4 and (54)--(58).

\noindent
3) For $s \in \mathbb{N}\,$
\eq
\sum_{\la=0}^{\la =l-1}\frac{1}{2^{\la}\,\la!}\
\int_\La^{\Lao} d\La'\ \La'^{-s-1} \
\Bigl(\,\log^{\la}(\sup({|\vec{p}|_{2n} \over \ka'},{\ka'\over m}))
\,+\,   [\la!]^{1/2}\,\Bigr)
\label{log}
\eqe
\[
\ \le\
5\ \frac{\La^{-s}}{s}\sum_{\la=0}^{\la =l-1} \frac{1}{2^{\la} \ \la!}\,
\log^{\la}\sup({|\vec{p}|_{2n} \over \ka},{\ka \over m})\ .
\]
We wrote $\ka' = \sup(\La ',m)\,$.
For the proof\footnote{In fact, the proof in \cite{Ko}
is given for $\La \ge m$, but it can be extended to  $\La < m$
without any problem.} see Lemma~5 in~\cite{Ko}.
\\

\noindent
4) For integers
$n\,,\,\,n_1\,,\, n_2 \ge 1\,$,
$\ l\,,\, l_1\,,\, \la_1\,,\ l_2\,,\, \la_2 \ge 0\,$
\[
\sum_{\begin{array}{c}_{l_1+l_2=l},\atop
_{n_1+n_2=n+1},\\_{\la_1 \le l_1, \,\la_2 \le l_2},
\atop _{\la_1+\la_2=\la}
 \end{array} }
\frac{1}{(l_1+1)^2\, (l_2+1)^2\,n_1^2\,n_2^2}\
\frac{n!}{ n_1!\,n_2!}\
\frac{\la!}{\la_1!\,\la_2!} \
 \frac{(n_1+l_1-1)!\ (n_2+l_2-1)!}{(n+l-1)!}
\]
\eq
\le\
20\ \frac{1}{(l+1)^2}\ \frac{1}{n^2}\  .
\label{1}
\eqe
For the proof see Lemma~2 in~\cite{Ko}.
\\

\noindent
5) We will repeatedly
use  bounds on the Hermite polynomials
$\,H_n(x) = (-1)^n\, \e ^{x^2}\  \frac{d}{d x^n} \e ^{-x^2} \,$:
\eq
H_n(x) \, \le \,
k\ \sqrt{n!}\ 2^{n/2}\  \e^{x^2/2}\ , \quad k=1.086\ldots\ .
\label{Sa}
\eqe
For a proof see~\cite{Sa},~p.~324. It then follows directly from this bound
that
\eq
\Bigl| \pa ^{w} \e^{-\frac{q^2+m^2}{\La ^2}} | \ \le\
k\ \La ^{-|w|}\ \sqrt{|w|!}\ 2^{\frac{|w|}{2}}\ \e^{-\frac{q^2}{2\La ^2}}\
 \e^{-\frac{m^2}{\La ^2}} \ .
\label{w}
\eqe

\subsection{Bounds on higher derivatives of CAG's without insertions}

Bounds on higher derivatives of CAG's
are proven inductively with the aid of the flow equation.
As compared to the bounds to be found in the literature~\cite{Ko}
the new ingredient here is a sufficiently precise control of those
bounds as regards their dependence on the the number of derivatives
$|w|$.
We want to show

\begin{prop}
\label{propwithout}
There exists a constant $\,K>0\,$ such that for $2n+|w|\ge 5\,$
\eq
|\pa^{w} {\cal L}_{2n,l}^{\La,\Lao}(p_1,\ldots,p_{n-1})|\leq \sqrt{|w|!}
\,\La^{4-2n-|w|}\
K^{(2n+4l-4)(|w|+1)}\ (n+l-2)! \
\sum_{\la=0}^{\la =\ell(n,l)}
\frac{\log^{\la}(\sup(\frac{|\vec{p}|}{\kappa},\frac{\ka}{m}))}{2^{\la}\,\la!}
\ .
\label{propout}
\eqe
Here
\eq
\ell(n,l) = l\ \mbox{ if }\ n\ge 2\ ,
\quad
\ell(n,l) = l-1\ \mbox{ if }\  n=1\ .
\label{til}
\eqe
\end{prop}

\noindent
The proposition is a consequence of the subsequent

\begin{lemma}\label{lablemma1}
There exists a constant $\,K>0\,$ such that for $2n+|w|\ge 5\,$
\eq
  |\pa^{w} {\cal L}_{2n,l}^{\La,\Lao}(p_1, \ldots, p_{2n-1})|
\leq \sqrt{|w|!}
\ \La^{4-2n-|w|}\
 \frac{K^{(2n+4l-4)(|w|+1)}}{(l+1)^2\, n!\,n^3}\ (n+l-1)! \
\sum_{\la=0}^{\la =\ell}
\frac{\log^{\la}(\sup(\frac{|\vec{p}|}{\kappa},
\frac{\ka}{m}))}{2^{\la}\,\la!}\ .
\label{lemma1}
\eqe
\end{lemma}

\noindent
{\sl Remark~:}
The Lemma is sharper than the proposition, and the stated bound is
suited as an inductive statement for its proof. Subsequently
we will however use the (shorter) bound from the proposition.

\noindent
{\it Proof~:}\\
The proof is based on the standard inductive scheme
which goes up in $n+l\,$ and for given  $n+l\,$ goes up in $l$,
and for given $n,\,l\,$ descends in $|w|\,$.
For $2n+|w| \le 4\,$ we will use the bounds
from the Theorem and Proposition\footnote{We have slightly simplified
the respective expressions which is possible if admitting for a
slightly larger $K\,$ as compared to \cite{Ko}.} in \cite{Ko}~:
\eq
  | {\cal L}_{4,l}^{\La,\Lao}(\vec{p})|\ \leq\
\frac{K^{2l}}{(l+1)^2\, 2^4}
\ (1+l)!\, \sum_{\la=0}^{\la =l}
\frac{\log^{\la}\bigr(\sup({|\vec{p}|\over \ka},{\ka\over
    m})\bigl)}{2^{\la}\,\la!}\
\ ,
\label{prop40}
\eqe
\eq
  |\pa^{w} {\cal L}_{2,l}^{\La,\Lao}(p)|\leq
\sup(|p|,\ka)^{2-|{w}|}\ \frac{K^{2l-1}}{(l+1)^2}
\ l!\,\sum_{\la=0}^{\la =l-1}
\frac{\log^{\la}\bigr(\sup({|p| \over \ka},{\ka\over m})\bigl)}{2^{\la}\,\la!}
\ .
\label{prop20}
\eqe
\\

\noindent
A) {\it The first term on the r.h.s. of the FE}\\[.1cm]
Integrating the FE (\ref{feq0}) w.r.t. the flow
parameter $\La'$ from $\La\,$ to $\Lao\,$ gives the
following bound\footnote{We assume $\,l \ge 1\,$, otherwise
the contribution is zero.}
 for the  first term on the r.h.s. of the FE
(writing $\ka' =\sup(\La',m)\,$):
\[
\int_\La^{\Lao} d\La' \int_k {2 \over \La'^3}\
\e^{-{{k^2+m^2}\over \La'^2}}\
\ \La'^{4-(2n+2)-|w|}\ \sum_{\la=0}^{\la =l-1}
\frac{\log^{\la}(\sup (\frac{|\vec p|_{2n+2}}{\ka'},{\ka'\over m}))}
{2^{\la}\,\la!}
\]
\[
\times\,\frac{(2n+1)(2n+2)}{2}\ \sqrt{|w|!}\
\frac{K^{(2n+4l-6)(|w|+1)}}{l^2\, (n+1)!\,(n+1)^3}\, (n+l-1)!\
\]
\eq
\leq \  (\frac{n}{n+1})^3\ (2n+1)\
\frac{K^{(2n+4l-6)(|w|+1)}}{l^2\, n!\,n^3} \
(n+l-1)!\
\sqrt{|w|!}\ \sum_{\la=0}^{\la =l-1}
\frac{1}{2^{\la}\,\la!}
\label{1stin1}
\eqe
\[
\times\
\int_\La^{\Lao} d\La'\ \La'^{3-2n-|w|} \  \e^{-{m^2 \over {\La'^2}}}\
\int_k \  \e^{-{k^2 \over {\La'^2}}}\
\log^{\la}
(\sup (\frac{|\vec{p}|_{2n+2}}{\kappa'},{\ka'\over m}))
\ .
\]
Using
\ben
|\vec{p}|_{2n+2} \ \le \
|\vec{p}|+|k|
\een\label{2n2n}
we bound the momentum integral with the aid of (\ref{momint}).
On performing the integral over $\La'$ in (\ref{1stin1}) and using (\ref{log})
we therefore obtain the following bound for (\ref{1stin1})
\eq
 \La ^{4-2n-|w|}\
\frac{K^{(2n+4l-4)(|w|+1)}}{l^2\, n!\,n^3} \ (n+l-1)!\
\sqrt{|w|!}
\sum_{\la=0}^{\la =\ell}
\frac{\log^{\la}(\sup (\frac{|\vec{p}|}{\kappa},{\ka\over m}))}
{2^{\la}\,\la!}
\label{erg}
\eqe
\[
\times\
\Bigl[5\  K^{-2(|w|+1)}\ 2^{|w|}\Bigr]\
 (\frac{n}{n+1})^3\ \frac{2n+1}{2n+|w|-4}\ .
\]
We realize that (\ref{erg}) is smaller than the inductive bound
divided by 2 if $K\,\ge \, 4\,$.\\[.2cm]
B) {\it The second term on the r.h.s. of the FE}\\[.1cm]
We assume without loss that  $2n+4l \ge 6\,$, otherwise
the contribution is zero.
Subsequently we will also assume that neither term is a two-point function
with $|w| \le 1\,$. If one of them is so, we  first have to bound the term
$\sup(q,\ka')^{2-|{w}|}\,$ arising from (\ref{prop20}) together
with the exponential $\e^{-\frac{q^2}{2\La'^2}}\,$ from the differentiated
propagator,
remember also (\ref{Sa}, \ref{w}),  by $2\,\La'^{2-|{w}|}\,$.
Afterwards this contribution can be absorbed into the
subsequent proof at the cost of a factor of 2 in the lower bound on
$K\,$.\\[.1cm]
 Integrating the inductive bound on the
{\sl second} term on the r.h.s. of the FE from  $\La\,$ to $\Lao\,$
then gives us  the following bound - where we also understand that the
$\sup$ w.r.t. the permutations of the momentum attributions
has been taken
\[
\int_\La^{\Lao} d\La' \  \La'^{8-(2n+2)-|w_1|-|w_2|}
\ K^{(2n+4l-6)(|w_1|+|w_2|+2)}\!\!\!\!\!\!\!\!\!\!\!
\sum_{
{\tiny
\begin{array}{c}
l_1+l_2=l,\\
w_1+w_2+w_3=w,\\
n_1+n_2=n +1
\end{array}
}
}
\!\!\!\!\!\!\!\!\!\! 2\,c_{\{w_i\}}\
\frac{ n_1}{(l_1+1)^2\, n_1!\,n_1^3} \ \frac{ n_2}{(l_2+1)^2\, n_2!\,n_2^3}
\]
\[
\times\
\sqrt{|w_1|!}\
(n_1+l_1-1)!\sum_{\la_1=0}^{\la_1 =\ell_1}
\frac{\log^{\la_1}\bigr(\sup({|\vec{p}|
\over \ka '},{\ka '\over m})\bigl)}{2^{\la_1}\,\la_1!}\
\frac{2}{\La'^3}\ |\pa^{w_3} \  \e^{-\frac{q^2+m^2}{\La '^2}}|
\]
\[
\times\
\sqrt{|w_2|!}\
(n_2+l_2-1)!\ \sum_{\la_2=0}^{\la_2 =\ell_2}
\frac{\log^{\la_2}
\bigr(\sup({|\vec{p}|\over \ka '},{\ka'\over m})\bigl))}
{2^{\la_2}\,\la_2!}
\]
\[
\le
\sum_{
{\tiny
\begin{array}{c}
l_1+l_2=l,\\
n_1+n_2=n +1,\\
\la_1 \le l_1, \,\la_2 \le l_2
\end{array}
}
}
\frac{1}{(l_1+1)^2\, (l_2+1)^2}\ \frac{1}{n_1^2\,n_2^2}
\ \frac{n!}{ n_1!\,n_2!} \ \frac{(\la_1+\la_2)!}{\la_1!\,\la_2!}
\ \frac{(n_1+l_1-1)!\ (n_2+l_2-1)!}{(n+l-1)!}
\]
\[
\times\ 2\ K^{(2n+4l-6)(|w|+2)}\ \frac{(n+l-1)!}{n!}
 \int_\La^{\Lao} d\La' \  \La'^{6-2n-|w_1|-|w_2|}\
\frac{\log^{\la_1+\la_2}
\bigr(\sup({|\vec{p}|\over \ka '},{\ka '\over m})\bigl)}
{2^{\la_1+\la_2}\,(\la_1+\la_2)!}
\]
\[
\times
\sum_{
{\tiny
\begin{array}{c}
w_1+w_2+w_3=w
\end{array}
}
}
 c_{\{w_i\}}\ \frac{2}{\La'^3}\  |\pa^{w_3} \ \e ^{-\frac{q^2+m^2}{\La '^2}}|\
\sqrt{|w_1|!\, |w_2|!}\ .
\]
Using (\ref{1}, \ref{w})
we then arrive at the bound\footnote{note that if $2n=2$
we have $2n_1=2n_2=2\,$, and the restriction to
$\la \le \ell$ in the sum over $\la\,$ is justified.}
\[
20\ \frac{1}{(l+1)^2}\ \frac{1}{n^2}\ 2\ K^{(2n+4l-6)(|w|+2)}\
\frac{1}{n!}\ (n+l-1)!\int_\La^{\Lao} d\La' \  \La'^{3-2n-|w|}\
\sum_{0\le \la \le\ell} \frac{\log^{\la}
\bigr(\sup({|\vec{p}|\over \ka '},{\ka '\over m})\bigl)}{2^{\la}\,\la!}
\]
\eq
\times
\sum_{w_i}
 c_{\{w_i\}}\, 2^{\frac12 |w_3|}\ k\,
 \sqrt{|w_1|!\, |w_2|!\, |w_3|!}\ .
\label{2ir}
\eqe
Using also (\ref{log})
 we verify the inductive bound
\[
\La^{4-2n-|w|}\ K^{(2n+4l-4)(|w|+1)}\
\frac{1}{(l+1)^2}\ \frac{1}{n^3}
\frac{1}{n!}\ (n+l-1)!\ \sqrt{|w|!}
\sum_{0\le \la \le \ell} \frac{\log^{\la}
\bigr(\sup({|\vec{p}| \over \ka },{\ka \over m})\bigl)}{2^{\la}\,\la!}
\]
multiplied by $1/4$,
on imposing the lower bound on $K\,$
\eq
K^{-2(|w|+2)}\ 40\ \frac{n}{2n+|w|-4}\
\sum_{w_i} c_{\{w_i\}}\, 2^{\frac12 |w_3|}\
\le \ 1/4\ ,
\label{bdk3}
\eqe
which is satisfied if
\[
K \ge  (640)^{\frac14}\ \ 3^{\frac{1}{2}}\ .
\]
\qed

\noindent
The following variant of Lemma~\ref{lablemma1} is proven analogously~:\\
\begin{cor}\label{cor00}
There exists a constant $\,K>0\,$ such that for
$2n+|w| \ge 5\,$
\\[.1cm]
\bena
  |\pa^{w} {\cal L}_{2n,l}^{\La,\Lao}(\vec{p})| &\leq& \sqrt{|w|!\,
(|w|+2n-4)!}\,\ka^{4-2n-|w|}\
\frac{K^{(2n+4l-4)(|w|+1)}}{n!} \ (n+l-1)! \non\\
&& \times \ \
\sum_{\la=0}^{\la =l}
\frac{\log^{\la}(\sup(\frac{|\vec{p}|}{\kappa},\frac{\ka}{m}))}{2^{\la}\,\la!}
\ .
\label{corr}
\eena
As a consequence, the ``gradient expansion'' of the effective action
\ben\label{gradient}
L^{0,\infty}(\varphi) = \sum_{n,l,w}  \int_{\mr^4} a_{w,n,l} \
\varphi(x) \ \partial^{w_1} \varphi(x) \cdots \partial^{w_{n-1}}
\varphi(x) \ d^4 x
\een
with
\ben
a_{w,n,l} := \frac{\hbar^l}{w!}\ (-i\partial)^w \L^{0,\infty}_{n,l}(\vec 0)
\een
converges absolutely for each fixed loop order $l$, 
for each fixed $n$, and each Schwartz-space
configuration $\varphi$ such that $\hat \varphi(p)$ has support in  a
sufficiently small ball around $p=0$ in
momentum space. Furthermore, the expansion in $l$ is locally
Borel summable.
\end{cor}

The bound~\eqref{corr} is weaker than the  one  of
eq.~\eqref{lemma1}, in the sense
that it replaces $\, \sqrt{|w|!}\,$ by
$\,\sqrt{|w|!\,(|w|+2n-4)!}\,$,
and
stronger  in the sense
that it replaces  $\,  \La ^{4-2n-|w|}\,$  by
$\,  \ka ^{4-2n-|w|}\,$ (\ref{ka}).
In the proof there is no change as regards the first term
on the r.h.s. of the FE~; as regards the second term we  use
 the bound~(\ref{la4})\footnote{Note that in the previous proof the
factor of $\e^{-m^2/\La ^2}\,$ is simply bounded by one and thus is
still at our disposal.}  which permits to transform
negative powers of $\La$ into  negative powers of $\ka$
at the cost of a square root of a factorial.\\
In the following we will only need
Proposition~\ref{propwithout}.


\subsection{Bounds on CAG's with one insertion}
Throughout this section, we fix a monomial $\O_A$ with
$A = \{n',w'\}$, and we denote the
dimension of this monomial by
\ben
D' := n'+|w'| = [A]  \,\,
.
\een
For simplicity, we also assume that $n'$ is even,
the odd case can be treated similarly.
We begin by rewriting the FE~\eqref{fe1ins} with this insertion,
with additional momentum derivatives:
\eq
\pa_{\La} \pa^{w} \, {\cal L}^{\La,\Lao}_{2n,l,D}(\O_A;p_1,\ldots p_{2n}) =
\left( {2n+2 \atop 2} \right) \int_k {\dot C}^{\La}(k)\,\pa ^w
{\cal L}^{\La,\Lao}_{2n+2,l-1,D}(\O_A; k,-k, p_1,\ldots p_{2n})
\label{feq}
\eqe
\[
-\!\!\!\!\!\!\!\!\!\!\!\!
\sum_{
{\tiny
\begin{array}{c}
l_1+l_2=l,\\
w_1+w_2+w_3=w,\\
n_1+n_2=n +1
\end{array}
}
}
\!\!\!\!\!\!\!\! \!\!\!\!4 n_1\,n_2\,
c_{\{w_j\}} \mathbb{S} \Biggl[ \pa^{w_1} {\cal
  L}^{\La,\Lao}_{2n_1,l_1,D}(\O_A; q, p_1,\ldots,p_{2n_1-1})\,
\,\pa^{w_3} {\dot C}^{\La}(q)\,\,
\pa^{w_2}{\cal L}^{\La,\Lao}_{2n_2,l_2}(p_{2n_1},\ldots,p_{2n})
\Biggr]
\]
As always in this subsection, the insertion is at the point $x=0$.
Inspection of the FE shows that the renormalizability proof
for the functions  ${\cal L}^{\La,\Lao}_{n,l,D}\,$ can be performed
on using the same
inductive scheme as the one used for the  ${\cal L}^{\La,\Lao}_{n,l}\,$,
namely going up in $n+l$,  for fixed  $n+l$ ascending in
$l\,$, and for  fixed $n,\,l\,$   descending in $|w|\,$.
Bounds on the functions without insertions
${\cal L}^{\La,\Lao}_{n,l}\,$ are taken from the previous section.
The boundary conditions for the
${\cal L}^{\La,\Lao}_{n,l,D}\,$
were given above in  eqs.~\eqref{br11},~\eqref{br21}.
We consider the case $D=D'=\,n'+|w'|\,$,  (\ref{br2}), and denote
${\cal   L}^{\La,\Lao}_{2n,l,D}(\O_A; \vec  p)\ $
simply by
$\ {\cal   L}^{\La,\Lao}_{2n,l}(\O_A; \vec  p)\ $
if $\ D = [A]\,$.

\noindent
\begin{thm}\label{thm2}
There exists a constant $\,K>0\,$ such that for $\La > 0\,$
\[
  |\pa^{w} {\cal L}_{2n,l}^{\La,\Lao}(\O_A;\vec{p})|
\leq\La ^{D-2n-|w|}\,
 K^{(4n+8l-4)|w|}\
 K^{D(n+2l)^3 }\
\]
\eq
\times\
\sqrt{|w'|!\ |w|!}\sum_{\mu=0}^{d(D,n,l,w)}\frac{1}{\sqrt{\mu!}}\
 (\frac{|\vec p|}{\La})^{\mu}\,
\sum_{\la=0}^{\ell'(n,l)}
\frac{\log^{\la}(\sup(\frac{|\vec{p}|}{\ka},\frac{\ka}{m}))}{2^{\la}\,\la!}\ .
\label{theo1}
\eqe
We set
\eq
d(D,n,l,w)~:=  D(2n+2l)+ \sup(D+1-2n-|w|,\,0)\ ,
\label{d}
\eqe
\eq
\ell'(n,l)~:= 2l+n-1\ .
\label{l'}
\eqe
\end{thm}

\noindent
{\it Proof~:}\\[.1cm]
We use the  notation and the bounds of
Proposition \ref{propwithout}  and proceed similarly as there.
If not written explicitly the arguments of $\ell'$ are supposed
to be $n,\,l\,$, those of $\ell_1'$ to be  $n_1,\,l_1\,$.
We start considering\\[.1cm]
{\it I)  Irrelevant terms  with  $2n+|w| >D\,$~:\\
A) \it The first term on the r.h.s. of the FE}\\[.1cm]
Integrating the inductive bound on the
{\sl first} term from the r.h.s. of the FE (\ref{feq}) over  $\La'\,$
between $\Lao\,$ and  $\,\La \,$
gives  the following bound\footnote{Assuming $l\ge 1$, otherwise the
  contribution is zero, and writing $\ka ' = \sup(\La ',m)\,$.}:
\[
\int_\La^{\Lao} d\La' \int_k {2 \over \La'^3}\
\e^{-{{k^2+m^2}\over \La'^2}}\
\ \La'^{D-(2n+2)-|w|}\
 \sum_{\mu=0}^{d(D,n+1,l-1,w)} \frac{1}{\sqrt{\mu!}}\
(\frac{|\vec p|_{2n+2}}{\La'})^{\mu}\,
\sum_{\la=0}^{\la =\ell'-1}
\frac{\log^{\la}(\sup (\frac{|\vec p|_{2n+2}}{\ka'},{\ka'\over m}))}
{2^{\la}\,\la!}
\]
\[
\times\,\left( {2n+2 \atop 2} \right)
\sqrt{|w'|!\ |w|!}\ K^{(4n+8l-8)|w|}\ K^{D(n+2l-1)^3 }\
\]
\[
\leq \ \left( {2n+2 \atop 2} \right) \ 
\sqrt{|w'|!\ |w|!}\ K^{(4n+8l-8)|w|}\ K^{D(n+2l-1)^3}\
\sum_{\la=0}^{\la =\ell'-1}
\frac{1}{2^{\la}\,\la!}
\]
\eq
\times\
\int_\La^{\Lao} d\La'\ \La'^{D-1-2n-|w|} \
 \sum_{\mu=0}^{d(D,n,l,w)} \frac{1}{\sqrt{\mu!}}
\int_{k/\La'}
(\frac{|\vec p|_{2n+2}}{\La'})^{\mu}\,
\log^{\la}
(\frac{|\vec{p}|_{2n+2}}{\ka'},{\ka'\over m})
\ \e^{-{k^2\over {\La'^2}}}\ .
\label{stin}
\eqe
Using  (\ref{pdef}) we show that
\[
|\vec{p}|_{2n+2} \ \le \
|\vec{p}|+|k|
\]
and bound the momentum integral by
\[
\sum_{\mu=0}^{d}\frac{1}{\sqrt{\mu!}}\
\sup_x\{ \e^{-{x^2/2}}\ (\frac{|\vec p|_{2n+2}}{\La'})^{\mu}\}\
\int_x \
 \e^{-{x^2/2}}\
\log^{\la}(\sup
(\frac{|\vec{p}|_{2n+2}}{\ka'},{\ka'\over m}))
\qquad (x =\frac{k}{\La' })
\]
\eq
\le\
\Bigl[\sum_{\mu=0}^{d}\frac{1}{\sqrt{\mu!}}\
\sum_{\rho=0}^{\mu} \left({\mu \atop \rho}\right)
\ (\frac{|\vec{p}|}{ \La'})^{\rho}\
2^{\frac{\mu-\rho}{2}}\ (\frac{\mu-\rho}{2})!\Bigr]\
\Bigl[  \,\log^{\la}(\sup
(\frac{|\vec{p}|_{2n}}{\ka'},{\ka'\over m}))\,+\, (\la!)^{1/2} \Bigr]
\label{sep2}
\eqe
with the aid of (\ref{momint}).
The first factor in (\ref{sep2}) can then be bounded by
\eq\label{sep21}
\sum_{\mu=0}^{d}\frac{1}{\sqrt{\mu!}}\
\sum_{\rho=0}^{\mu} \left({\mu \atop \rho}\right)
\, (\frac{|\vec{p}|}{ \La'})^{\rho}\
(\frac{\mu-\rho}{2})! \ 2^{\frac{\mu-\rho}{2}}
\ \le\
\sum_{\rho=0}^{d}  (\frac{|\vec{p}|}{ \La'})^{\rho}\
\sum_{\mu=\rho}^{d}\frac{1}{\sqrt{\mu!}}\ \left( {\mu \atop \rho}\right) \
(\frac{\mu-\rho}{2})! \ 2^{\frac{\mu-\rho}{2}}
\eqe
\[
\le\ \sum_{\mu=0}^{d} \frac{1}{\sqrt{\mu!}}\
 (\frac{|\vec{p}|}{ \La'})^{\mu}\
\sum_{\rho=0}^{d-\mu} \left({\rho+\mu \atop \mu} \right) \
(\frac{\rho}{2})! \ 2^{\frac{\rho}{2}}\
\sqrt{\frac{\mu!}{(\rho+\mu)!}}
\ \le \
2^{d}\ \sum_{\mu=0}^{d} \frac{1}{\sqrt{\mu!}}
 (\frac{|\vec{p}|}{ \La'})^{\mu}\ ,
\]
where the last bound is obtained from Stirling and binomial
type estimates.

\noindent
Performing the integral over $\La'$ in (\ref{stin}) and using
(\ref{log}),
we therefore obtain the following bound for (\ref{stin}):
\[
 \La ^{D-2n-|w|}\
K^{D(n+2l)^3}\ K^{(4n+8l-4)|w|}\  \
\sum_{\mu=0}^{d} \frac{1}{\sqrt{\mu!}}\
 (\frac{|\vec{p}|}{ \La})^{\mu}\
\sum_{\la=0}^{\la =\ell'-1}
\frac{\log^{\la}(\sup(\frac{|\vec{p}|}{ \ka},{\ka\over m}))}{2^{\la}\,\la!}
\]
\eq
\times\
K^{-D[(n+2l)(n+2l-1)+1]-|w|}\
\Bigl[ 5\frac{(n+1) (2n+1)}{2n+|w|-D}\ 2^{d}
\ K^{-2D(n+2l)(n+2l-1)-3|w|}
\Bigr]
\ .
\label{erg1}
\eqe
As a consequence, this contribution to
$\pa^{w} {\cal L}_{2n,l}^{\La,\Lao}(\O_A,\vec{p})\,$
{\bf satisfies the inductive bound multiplied by }
\eq
 {\cal A}(D,n,l,w)~:= 1/8\ K^{-|w|} K^{-2D[(n+2l)(n+2l-1)+1]}\ ,
\label{res}
\eqe
if we assume $K\,$ to be sufficiently large
such that
\eq
5 (n+1) (2n+1)\ 2^{d}
\ K^{-2D(n+2l)(n+2l-1)-3|w|} \ \le \
1/8\ .
\label{K}
\eqe

\noindent
{\it B) The second term on the r.h.s. of the FE}\\[.1cm]
Integrating the inductive bound on the
{\sl second} term on the r.h.s. of the FE over  $\La'\,$
between   $\,\La\,$ and $\Lao\,$
gives  the following bound\footnote{\label{15}
Note that the lowest possible value of
$4n+8l-4$ which may give a nonvanishing contribution
on the r.h.s. is  $4$. This is realized for $(n=2, l=0)$.
Thus the corresponding exponent of $K$
in the inductive bound is never negative.
A negative exponent could give a bound incompatible with the boundary
contributions from (\ref{br2}).},
using that $|\vec p|_{2n_1},|\vec
p|_{2n_2} \le |\vec p|_{2n}\equiv |\vec
p|\,$ and  taking the
$\sup$ w.r.t. the permutations of the momentum assignments:
\[
\int_\La^{\Lao} d\La' \  \La'^{D+4-2n-2-|w_1|-|w_2|}
\ K^{(4n_1+8l_1-4+2n_2+4l_2-4)(|w_1|+|w_2|+1)}\
\ K^{D(n_1+2l_1)^3}\!\!\!\!\!\!\!\!
\]
\[
\times\!\!\!\!\!\!\!
\sum_{\begin{array}{c}\\[-.9cm]_{l_1+l_2=l,}\\[-.3cm]
_{w_1+w_2+w_3=w,}\\[-.3cm]
_{n_1+n_2=n+1} \end{array} }\!\!\!\!\!\!\!
 4\,c_{\{w_i\}}\ n_1 n_2\ \sqrt{
|w'|!\ |w_1|!}\
\sum_{\mu=0}^{d(D,n_1,l_1,w_1)}\frac{1}{\sqrt{\mu!}}\
 (\frac{|\vec p|}{\La'})^{\mu}\,
\sum_{\la_1=0}^{\la_1 =\ell_1'}
\frac{\log^{\la_1}\bigr(\sup({|\vec{p}|
\over \ka '},{\ka '\over m})\bigl)}{2^{\la_1}\,\la_1!}\
\]
\[
\times\
\frac{2}{\La'^3}\ |\pa^{w_3} \  e ^{-\frac{q^2+m^2}{\La '^2}}|\
\sqrt{|w_2|!}\
(n_2 + l_2- 2)!\
\sum_{\la_2=0}^{\la_2 ={\ell}_2}
\frac{\log^{\la_2}
\bigr(\sup({|\vec{p}|\over \ka '},{\ka'\over m})\bigl)}
{2^{\la_2}\,\la_2!}
\]
\[
\le
\sum_{\begin{array}{c}\\[-.9cm]_{l_1+l_2=l,}\\[-.3cm]
_{n_1+n_2=n+1,} \\[-.2cm]_{\la_1 \le \ell'_1, \,\la_2 \le \ell_2}\end{array} }
\!\!\!\!\!(n_2 + l_2)!\
4\, n_1\,\, \frac{(\la_1+\la_2)!}{\la_1!\,\la_2!}
\  K^{(4n+8l-6)(|w|+1)}\ K^{D(n_1+2l_1)^3}
\]
\[
\times\
 \int_\La^{\Lao} d\La' \  \La'^{D+2-2n-|w_1|-|w_2|}\
\sum_{\mu=0}^{d(D,n_1,l_1,w_1)} \frac{1}{\sqrt{\mu!}}\
(\frac{|\vec p|}{\La'})^{\mu}\,
\frac{\log^{\la_1+\la_2}
\bigr(\sup({|\vec{p}|\over \ka '},{\ka '\over m})\bigl)}
{2^{\la_1+\la_2}\,(\la_1+\la_2)!}
\]
\[
\times
\sum_{w_1+w_2+w_3=w}
c_{\{w_i\}}\ \frac{2}{\La'^3}\  |\pa^{w_3}
\ \e ^{-\frac{q^2+m^2}{\La '^2}}|
\
|w'|!\ |w_1|!\ |w_2|!\ .
\]
Using (\ref{w}) and the fact that $\ell'_1+{\ell}_2 \le \ell'\, $,
we  arrive at the bound
\[
\sum_{\begin{array}{c}\\[-.9cm]_{l_1+l_2=l},\\[-.3cm]
_{n_1+n_2=n+1}
 \end{array} }\!\!\!\!
(n_2+l_2)!\, 4\,n_1\ \ell'\  2^{\ell'}\
\ K^{(4n+8l-6)(|w|+1)}\ K^{D(n_1+2l_1)^3}
\sum_{w_i}
 c_{\{w_i\}}\, 2^{\frac12 |w_3|}\ k
\sqrt{
|w'|!\, |w_3|!\, |w_1|!\, |w_2|!}
\]
\eq
\times\
\sum_{\mu=0}^{d(D,n_1,l_1,w_1)}\frac{1}{\sqrt{\mu!}}\
 (\frac{|\vec p|}{\La})^{\mu}
\int_\La^{\Lao} d\La' \  \La'^{D-2n-|w|-1}\
\sum_{0\le \la \le \ell'} \frac{\log^{\la}
\bigr(\sup({|\vec{p}|\over \ka '},{\ka '\over m})\bigl)}{2^{\la}\,\la!}\ .
\label{2ir2}
\eqe
Using also (\ref{log}) we verify the  bound
\eq
\La^{D-2n-|w|}\ K^{(4n+8l-4)|w|+D(n+2l)^3}\
\sqrt{
|w'|!\ |w|!}
\sum_{\nu=0}^{d(D,n_1,l_1,w_1)} \!\!\!\! \frac{1}{\sqrt{\mu!}}\
(\frac{|\vec p|}{\La})^{\mu}
\sum_{0\le \la \le \ell'}
 \frac{\log^{\la}
\bigr(\sup({|\vec{p}| \over \ka },{\ka \over m})\bigl)}{2^{\la}\,\la!}\ ,
\label{prov}
\eqe
 multiplied by (\ref{res})--on imposing the lower bound on $K\,$
\eq
K^{-2D(n+2l)[(n+2l)-1]+(4n+8l-6)-|w|}\, 5k \!\!\!\!\!\!\!\!\!\!
\sum_{\begin{array}{c}\\[-.9cm]_{l_1+l_2=l},\\[-.3cm]
_{n_1+n_2=n+1}
 \end{array} } \!\!\!\!\!\!
(n_2+l_2)!\,  4\,n_1\, \ell'\ 2^{\ell'}\!
\sum_{w_i}
c_{\{w_i\}}\, 2^{\frac{ |w_3|}{2}}
  \le
\frac18
\label{bdk}
\eqe
where we used that $\, n_1+2l_1 \le n+2l-1\,$.
Noting also $\, 2n_1+2l_1 \le 2n+2l-2\,$ we verify that
\eq
d(D,n_1,l_1,w_1) \le
d(D,n,l,w)\ ,
\label{d1}
\eqe
with the aid of definition (\ref{d}), so that
(\ref{prov}) is bounded by (\ref{theo1}), as required.\\[.1cm]
Adding the bounds on the first and second terms on the r.h.s. of the FE
we verify  the bound (\ref{theo1}) multiplied by $2   {\cal
  A}(D,n,l,w)\,$
for
$K\,$ sufficiently large to satisfy (\ref{K}), (\ref{bdk}).\\[.2cm]
\noindent
{\it II) Relevant terms at vanishing external momentum}\\[.1cm]
Relevant terms - i.e. $2n+|w|\le D\,$ -
are first constructed at zero external momentum
with the aid of the Taylor series
\eq
\pa_{\vec p }^v\ f_{2n}(\vec p)\,=\!\!\!\!
\!\!\sum_{|w| \le D-2n-|v|}\! \frac{{\vec p}^{\,w}}{w!}\,
[\pa_{\vec p }^{w+v} f_{2n}](0)\, +\!\!\!
\!\!\!\!\sum_{|w|=D+1-2n-|v|}\!\!\!\!\!\! {\vec p}^{\,w}
\int_0^1 d\tau  \frac{(1-\tau)^{|w|-1}}{(|w|-1)!}\,
[\pa_{\vec p }^{\,w+v} f_{2n}](\tau {\vec p})\ .
\label{sloe}
\eqe
We note that for the relevant terms we also have to take into account
the contribution from the boundary condition, see
 (\ref{br2}); the factor of $w!\,\delta_{w,w'}$
in the boundary condition  exhausts the factor of
$\sqrt{w!\, w'!}\,$ present in the inductive bound (\ref{theo1}),
which thus cannot be sharpened in this respect.
We  consider the r.h.s. of the FE for the term
$\,\pa_{\vec p }^{w+v}{\cal L}^{\La,\Lao}_{2n,l}(\O_A,
\vec 0)\,$ with $2n +|w+v| \le  D\,$. \\[.2cm]
A) {\it The first term on the r.h.s. of the FE}\\[.1cm]
Integrating the FE (\ref{feq}) w.r.t. $\La'$ from $0\,$
to $\La\,$--assuming again without loss of
generality $ l\ge 1$--gives the
following bound for the  first term on the r.h.s. of the FE:
\[
 \left({ 2n+2 \atop 2} \right)\
K^{(4n+8l-8)|w+v|} \, K^{D(n+2l-1)^3}\ \sqrt{|w'|!\, |w+v|!}\
\int_0^{\La} d\La'\  \La'^{D-(2n+2)-|w+v|}\ \frac{2}{\La'^3}
 \]
\[
\times
\int_k
\e^{-{{k^2+m^2}\over \La'^2}}
\sum_{\mu=0}^{d(D,n,l,w+v)} \frac{1}{\sqrt{\mu!}}
(\frac{|k|}{\La'})^{\mu}\,\sum_{\la=0}^{\la =\ell'-1}
\frac{\log^{\la}(\sup(\frac{|k|}{\ka'},\frac{\ka '}{m}))}
{2^{\la}\,\la!}
\]
\[
\leq \  \left( { 2n+2 \atop 2} \right)\
K^{(4n+8l-8)|w+v|}\ K^{D(n+2l-1)^3}\
\sqrt{|w'|!\ |w+v|!}\
\sum_{\mu=0}^{d(D,n,l,w+v)}
\sum_{\la=0}^{\la =\ell'-1}
\frac{1}{2^{\la}\,\la!}
\]
\eq
\times \ 2
\int_0^{\La} d\La' \ \La'^{D-(2n+1)-|w+v|}\
\e^{-{{m^2}\over \La'^2}}
 \
\int_{\frac{|k|}{\La'}}\frac{1}{\sqrt{\mu!}}
 (\frac{|k|}{\La'})^{\mu}\
\log^{\la}(\sup(\frac{|k|}{\ka'},\frac{\ka '}{m}))
\ \e^{-{k^2\over {\La'^2}}}\ .
\label{stin2}
\eqe
We bound the momentum integral as before in (\ref{sep2}) by
\eq
2^{\frac{\mu}{2}}\ \frac{1}{\sqrt{\mu!}}\
(\frac{\mu}{2})!\
[\log^\la (\frac{\ka'}{m}) \,+\,(\la!)^{1/2}] \ .
\label{sepp}
\eqe
Summing  over $\mu\,$, and using $(\frac{\mu}{2})!
\le 2^{-\mu/2}\, \sqrt{\mu+1}
\, \sqrt{\mu!}\,$,
the first term from (\ref{sepp}) can then be bounded by
\eq
\sum_{\mu=0}^{d(D,n,l,w+v)}
2^{\frac{\mu}{2}}\ 2^{-\frac{\mu}{2}}\ \sqrt{\mu+1}\
\le\
2\,d(D,n,l,w+v)^{3/2}\ .
\label{d2}
\eqe
Using that
\bena\label{respint}
&&\int_0^{\La} d\La'\ \La'^{D-(2n+1)-|w+v|}\ \log^\la (\frac{\ka'}{m})
\e^{-{{m^2}\over \La'^2}} \non \\
&&\le   \La^{D-2n-|w+v|}
\begin{cases}
\log^\la(\frac{\ka}{m}) & \text{if $D-2n-|w+v|>0$,}\\
2 (\la+1)^{-1}\log^{\la+1}(\frac{\ka}{m}) & \text{if $D-2n-|w+v|=0$,}
\end{cases}
\eena
we therefore obtain for (\ref{stin2}) the  bound
\bena
&&\hspace{2cm}\left( { 2n+2 \atop 2} \right)\
K^{(4n+8l-8)|w+v|}\ K^{D(n+2l-1)^3}\
\sqrt{|w'|!\ |w+v|!}\non\\
&&\hspace{2cm}
\times\ \La^{D-2n-|w+v|}\
2\,d(D,n,l,w+v)^{3/2}\
6 \sum_{\la=0}^{\la =\ell'}
\frac{1}{2^{\la} \la!}
\log^{\la} (\frac{\ka}{m})\ .
\label{2in}
\eena
As a consequence, this contribution to
$\pa^{w+v} {\cal L}_{2n,l}^{\La,\Lao}({\cal O}_A,\vec{0})\,$
 satisfies the inductive bound multiplied by (\ref{res})
(with $w \to w+v$)
under the condition that
\[
12 \ \left({ 2n+2 \atop 2} \right)\ d(D,n,l,w+v)^{3/2}\
  K^{-2D(n+2l)(n+2l-1)-3|w+v|}\
\le\ 1/8\ .
\]

\noindent
B) {\it The second term on the r.h.s. of the FE}\\[.1cm]
Integrating the inductive bound on the
{\sl second} term on the r.h.s. of the FE from  $0\,$ to $\La\,$
gives the following bound at zero momentum
\[
\sum_{\begin{array}{c}\\[-.9cm]_{l_1+l_2=l,}\\[-.3cm]
_{w_1+w_2+w_3=w+v,}\\[-.3cm]
_{n_1+n_2=n+1} \end{array} }\!\!\!\!\!\!\!\!\!\!\!\!
4\, n_1 n_2\,\int_0^\La d\La' \  \La'^{D+4-2n-2-|w_1|-|w_2|}
\ K^{(4n_1+8l_1-4+2n_2+4l_2-4)(|w_1|+|w_2|+1)}\
\ K^{D(n_1+2l_1)^3}
\]
\[
\times\
c_{\{w_i\}}\,
\sqrt{|w'|!\ |w_1|!}
\sum_{\la_1=0}^{\la_1 =\ell_1'}
\frac{\log^{\la_1}({\ka '\over m})}{2^{\la_1}\,\la_1!}\
\frac{2}{\La'^3}\
\bigg|\pa^{w_3} \  \e^{-\frac{q^2+m^2}{\La '^2}}\bigg|_{q=0}\
\sqrt{|w_2|!}\
(n_2 + l_2- 2)!
\sum_{\la_2=0}^{\la_2 =\ell_2}
\frac{\log^{\la_2}({\ka'\over m})}
{2^{\la_2}\,\la_2!}
\]
\[
\le
\sum_{\begin{array}{c}\\[-.9cm]_{l_1+l_2=l,}\\[-.3cm]
_{n_1+n_2=n+1,} \\[-.2cm]_{\la_1 \le \ell'_1, \,\la_2 \le \ell_2}\end{array} }
\!\!\!\!\!(n_2 + l_2)!\
4\, n_1\,\, \frac{(\la_1+\la_2)!}{\la_1!\,\la_2!}
\  K^{(4n+8l-6)(|w|+|v|+1)}\ K^{D(n_1+2l_1)^3}
\!\!\!\!\sum_{\begin{array}{c}\\[-.9cm]
_{w_1+w_2+w_3\atop
=w+v}\end{array} }\!\!\!\!
c_{\{w_i\}}
\]
\[
\times
\int_0^\La d\La' \  \La'^{D+2-2n-|w_1|-|w_2|}\
\frac{\log^{\la_1+\la_2}
({\ka '\over m})}
{2^{\la_1+\la_2}\,(\la_1+\la_2)!}\,
\frac{2}{\La'^3}\
\bigg|\pa^{w_3} \  \e^{-\frac{q^2+m^2}{\La '^2}}\bigg|_{q=0}
\,
\sqrt{|w'|! |w_1|! |w_2|!}\ ,
\]
remembering (\ref{l'}) and (\ref{til}) which
imply that $\ell_1'+ \ell_2 \le \ell'\,$.
Using (\ref{w})
we  obtain the bound
\[
\!\!\!\!\!\sum_{\begin{array}{c}\\[-.9cm]_{l_1+l_2=l,}\\[-.3cm]
_{n_1+n_2=n+1}
 \end{array} }\!\!\!\!\!\!\!\!
(n_2+l_2)!\, 4\,n_1\, l  2^l
\ K^{(4n+8l-6)(|w+v|+1)}\ K^{D(n_1+2l_1)^3}
\sum_{w_i}
 c_{\{w_i\}}\, k\, 2^{\frac12 |w_3|}\
\sqrt{|w'|!\, |w_3|!\, |w_1|!\, |w_2|!}
\]
\eq
\times\
\int_0^\La d\La' \  \La'^{D-2n-|w+v|-1}\ e^{-\frac{m^2}{{\La'} ^2}}\
\sum_{0\le \la \le \ell'} \frac{\log^{\la}
({\ka '\over m})}{2^{\la}\,\la!}\ .
\label{2r2}
\eqe
Using also (\ref{log})
and proceeding as in (\ref{respint}, \ref{2in})
we verify the inductive bound (\ref{theo1})
\[
\La^{D-2n-|w+v|}\ K^{(4n+8l-4)|w+v|+D(n+2l)^3}\
\sqrt{|w'|!\ |w+v|!}
\sum_{0\le \la \le \ell'}
 \frac{\log^{\la}({\ka \over m})}{2^{\la}\,\la!}\ ,
\]
 multiplied by (\ref{res}) (with $w \to w+v$)
on imposing the lower bound on $K\,$
\eq
6\,
K^{-2D(n+2l)(n+2l-1)+(4n+8l-6)-|w+v|}\, 5k \!\!\!\!\!\!\!\!\!\!
\sum_{\begin{array}{c}\\[-.9cm]_{l_1+l_2=l,}\\[-.3cm]
_{n_1+n_2=n+1}
 \end{array} } \!\!\!\!\!\!
(n_2+l_2)!\,  4\,n_1\, \ell'\ 2^{\ell'}\!
\sum_{\sum w_i =w+v}
c_{\{w_i\}}\, 2^{\frac{ |w_3|}{2}}
  \le
\frac18 \ .
\label{bdkr}
\eqe

For $\,|w+v| + 2n = D\,$
we have to add the boundary term from (\ref{br2}).
Since it is non-vanishing only if $w+v= w'$ we can bound it
by $\,\sqrt{ |w+v|! |w'|!}\,$ and it is thus accommodated for
by the bound from Theorem \ref{thm2} multiplied by  $\frac18\ $,
remember also footnote \ref{15}.\\[.2cm]
\noindent
{\it III)  Schwinger functions with $|v|+2n \le D\,$ at arbitrary
  external momenta}\\[.1cm]
We have to sum the series
(\ref{sloe}).
\[
|\pa_{\vec p }^v\,{\cal L}_{2n,l}^{\La,\Lao}(\O_A,\vec p)|\,=\,
\]
\eq
\Bigl|\! \!\! \!\sum_{|w| \le D-2n-|v|}\! \! \frac{{\vec p}^{\,w}}{w!}\,
\,\pa_{\vec p }^{w+v} {\cal L}_{2n,l}^{\La,\Lao}(\O_A,{\vec 0})
\,\, +\! \!\! \!\! \!\sum_{|w|=D+1-2n-|v|} \! \!\! \!
{\vec p}^{\,w} \int_0^1 \frac{(1-\tau)^{|w|-1}}{(|w|-1)!}\,
\,\pa_{\vec p }^{\,w+v}
{\cal L}_{1,2n,l}^{\La,\Lao}(\O_A,\tau {\vec p})\ \Bigr|
\label{dev}
\eqe
\eq
\leq\
\Bigg[
\sum_{|w| \le D-2n-|v|}  (\frac{\ |{\vec p}|}{\La})^{|w|}\
4\, {\cal A}(D,n,l,w+v)\ K^{(4n+8l-4)|w+v|}\
\frac{ \sqrt{|w'|!\ |w+v|!}}{w!}\
\sum_{\la=0}^{\la =\ell'}
\frac{\log^{\la}(\frac{\ka}{m})}{2^{\la}\,\la!}
\label{the}
\eqe
\[
+ \!\!\! \!\!\sum_{|w|=D+1-2n-|v|\,}\!\!\!
4\, {\cal A}(D,n,l,w+v)\ (\frac{|\vec p|}{\La})^{|w|}\
|w|\,K^{(4n+8l-4)|w+v|}\
\frac{\sqrt{|w'|!\,|w+v|!}}{|w|!}\
 \int_0^1 d\tau (1-\tau)^{|w|-1}
\]
\[
\times\sum_{\mu=0}^{d(D,n,l,w+v)} \frac{1}{\sqrt{\mu!}}
(\frac{\tau |\vec{p}|}{\La})^{\mu}\,
\sum_{\la=0}^{\la =\ell'}
\frac{\log^{\la}
(\sup(\frac{\tau |\vec{p}|}{\ka},\frac{\ka}{m}))}{2^{\la}\,\la!}
\ \Bigg]\
K^{D(n+2l)^3}\ \La^{D-2n -|v|}
\ .
\]
where ${\cal A}$ is given in (\ref{res}).
We used the induction hypothesis,
 after transforming powers of $p$ into powers
of $p$ over $\La$ multiplied by powers of $\La\,$.

\noindent
Using the following estimate
\[
\frac{ \sqrt{|w'|!\ |w+v|!}}{{w!}}\
\le\
\frac{\sqrt{|w'|!\ |v|!}}{\sqrt{|w|!}} \ 2^{\frac{|w+v|}{2}}\ \frac{|w|!}{w!}
\le\
\frac{\sqrt{|w'|!\ |v|!}}{\sqrt{|w|!}}
\ 2^{\frac{|w|+|v|}{2}}\ (8n)^{|w|} \ ,
\]
 we obtain  a bound for $2n+|v| \le  D$~:
\[
|\pa_{\vec p }^v\
{\cal L}_{2n,l}^{\La,\Lao}(\O_A,\vec p)|\leq
\La^{D-2n -|v|}\ \sqrt{|w'|!\ |v|!}\  K^{D(n+2l)^3 }
\]
\[
\times\ \Bigg[
\sum_{w,2n+|v+w| \le D} 4\, {\cal A}(D,n,l,w+v)\
\frac{1}{\sqrt{|w|!}}\ (\frac{\ |{\vec p}|}{\La})^{|w|}
\  2^{\frac{|w|+|v|}{2}}\ (8n)^{|w|} \
K^{(4n+8l-4)|w+v|}\ \sum_{\la=0}^{\la =\ell'}
\frac{\log^{\la}(\frac{\ka}{m})}{2^{\la}\,\la!}
 \]
\eq
+ \! \sum_{w,|w|=D+1-2n-|v|\,}\!\!\! 4\, {\cal A}(D,n,l,w+v)\
\frac{1}{\sqrt{|w|!}} \ (\frac{|\vec p|}{\La})^{|w|}\
|w|\,K^{(4n+8l-4)|w+v|}\
2^{\frac{|w|+|v|+2}{2}}\ (8n)^{|w|}
\eqe
\[
\times\ \int_0^1 d\tau (1-\tau)^{|w|-1}
\sum_{\mu=0}^{d(D,n,l,w+v)} \frac{1}{\sqrt{\mu!}}
(\frac{\tau |\vec{p}|}{\La})^{\mu}\,
\sum_{\la=0}^{\la =\ell'}
\frac{\log^{\la}
(\sup(\frac{ \tau |\vec{p}|}{\ka},\frac{\ka}{m}))}{2^{\la}\,\la!}
\Bigg]
\]
\[
\leq\
\La^{D-2n -|v|}\
\sqrt{|w'|!\ |v|!}\  \ K^{D(n+l)^3}\
\]
\[
\times\ \Bigg[
\sum_{w,2n+|v+w| \le D} 4\, {\cal A}(D,n,l,w) \
\frac{1}{\sqrt{|w|!}}\ (\frac{\ |{\vec p}|}{\La})^{|w|}
\  2^{\frac{|w|+|v|}{2}}\ (8n)^{|w|} \
K^{(4+8l-4)|w+v|}\ \sum_{\la=0}^{\la =\ell'}
\frac{\log^{\la}(\frac{\ka}{m})}{2^{\la}\,\la!}
 \]
\[
+ \!\!\!\!\!\!\!\! \sum_{|w|=D+1-2n-|v|\,}\!\!\!\!\!\!\!\!\!
4\, {\cal A}(D,n,l,w)\ |w|\, K^{(4n+8l-4)|w+v|}\,
2^{\frac{|w|+|v|+2}{2}}\ (8n)^{|w|}
\]
\eq
\times\
\sum_{\mu=0}^{d(D,n,l,w+v)+|w|} \frac{2^{\frac{\mu}{2}}}{\sqrt{\mu!}}
(\frac{ |\vec{p}|}{\La})^{\mu}\,
\sum_{\la=0}^{\la =\ell'}
\frac{\log^{\la}
(\sup(\frac{ |\vec{p}|}{\ka},\frac{\ka}{m}))}{2^{\la}\,\la!}
\Bigg]\ .
\eqe
These bounds are compatible with the induction hypothesis since\\
i)
\ben\label{ikey}
d(D,n,l,w+v)+|w| \le d(D,n,l,v)
\een
for $|w| \le D-2n-|v|+1\,$, as a consequence of the definition of
$d$ (\ref{d}),\\[.1cm]
ii)
\[
\sum_{|w| \le D+1-2n-|v|\,}
(8n)^{|w|}\ 2^{\frac{d(D,n,l,v)}{2}+|w|}\
|w|\,K^{(4n+8l-4)|w|} \ 4\, {\cal A}(D,n,l,w) \le\ 1
\]
for $K\,$ sufficiently large.\\[.1cm]

\qed
\\

\noindent
\begin{cor}\label{corr1}
For $\La \le m$ and $K$ sufficiently large we have the bounds
\[
  |\pa^{w} {\cal L}_{2n,l}^{\La,\Lao}(\O_A,\vec{p})| \
\leq
m^{D-2n -|w|}\
K^{(4n+8l-4)|w|}\
 K^{D(n+2l)^3}
\]
\eq
\times\ \sqrt{|w'|!\, |w|!\,[2n+|w| - D]_+!}
\sum_{\mu=0}^{d(D,n,l,w)}
 (\frac{|\vec p|}{m})^{\mu}\,
\sum_{\la=0}^{\la =\ell'}
\frac{\log_+^{\la}(\frac{|\vec{p}|}{m})}{2^{\la}\,\la!}\
\ .
\label{coreq}
\eqe
\end{cor}
\noindent
{\sl Remark:} These bounds show that the functions
$\,{\cal L}_{2n,l}^{\La,\Lao}(\O_A;\vec{p})\,$ have a
convergent Taylor expansion around zero momentum, since
the growth of the Taylor coefficients 
is bounded by $\ti K ^{|w|}\, |w|!$ for $w\,$ large
and suitable $\ti K\,$.\\[.1cm]
{\it Proof}~:
To prove the Corollary we may insert the bounds of
Theorem \ref{thm2} on the
r.h.s. of the FE. We may then use the factors
$\exp(-m^2/{\La'} ^2)\,$ present in both terms
to  bound the negative powers of $\Lambda\,$ by a square root of a
factorial~:
\[
\int_0^{m}d\La\,'\exp(-m^2/{\La'} ^2)\
\frac{\La'^{D-2n-|w|-\mu-1}}{\sqrt{\mu!}}
\,\le\,  m^{D-2n-|w|-\mu}\ \frac{\sqrt{(2n+|w|+\mu-D)_+!}}{\sqrt{\mu!}}
\]
\[
\le\ 2^{\frac12(2n+|w|+\mu-D)_+}\ m^{D-2n-|w|-\mu}\ \sqrt{(2n+|w|-D)_+!}\ .
\]
 These bounds cannot serve as
a viable induction hypothesis however, since the powers of momenta
(now without $\frac{1}{\sqrt{\mu!}}\,$)
would create additional square roots of
factorials in the next step of induction. \qed
\\

\noindent
For later use, we also note the
\begin{cor}\label{newcor}
The inductive proof of Theorem~\ref{thm2} is valid for a
somewhat larger constant $K$ also if
we replace $d$ in the statement of the theorem by $2d$,
or if we replace $\ell'$ by $\ell'+1$.
\end{cor}

{\it Proof}~: The key properties for $d$ that enter the proof are the
inequalities~\eqref{d1} and~\eqref{ikey}, which are evidently also
satisfied for $2d$. The key properties required for $\ell'=\ell'(n,l)$ are
that $\ell'(n+1,l-1) < \ell'(n,l)$ for $l \ge 1$ and
$\ell'(n_1,l_1)+\ell(n_2,l_2) < \ell'(n,l)$ on the r.h.s. of the FE,
where $\ell(n,l)$ is
as in eq.~\eqref{til}. These properties are
evidently also satisfied by the quantity $\ell'+1$. \qed


\subsection{Bounds on normal products with two insertions}

We now provide bounds on the normal products
$\L^{\La,\Lao}_{n,l,D}(\O_A
\otimes \O_B; \vec p)$ with two insertions.
Each of these insertions is a monomial in the basic fields with $A
= \{n',w'\}$ and $B = \{n'', w''\}$.
Again, we will assume for simplicity that both $n'$ and $n''$ are
even. We will use the notation $D'$
for the combined dimension of the two insertions,
\ben\label{inviewof}
D' := [A]+[B] = n'+n''+|w'|+|w''| \, .
\een
These normal products were defined above in sec.~\ref{sec2}
as the solutions to the FE's~\eqref{fe2ins}, and the boundary
conditions are given above in eq.~\eqref{br2a}
and eq.~\eqref{br2b}.
Our bounds are given in the following theorem:

\begin{thm}\label{thm3}
There exists a constant $\,K>0\,$ such that for $|w| \le D'+1$:
\[
  |\pa^{w} {\cal L}_{2n,l,D'}^{\La,\Lao}(\O_A \otimes \O_B;\vec{p})|
\leq \La^{D'-2n-|w|}\,
 K^{(4n+8l-4)|w|}\
 K^{D'(n+2l)^3 }\
\]
\eq
\sqrt{|w|! \ |w'|! \ |w''|!} \sum_{\mu=0}^{d'(n,l,w,D')}\frac{1}{\sqrt{\mu!}}\
 (\frac{|\vec p|}{\La})^{\mu}\,
\sum_{\la=0}^{\la =\ell'+1}
\frac{\log^{\la}(\sup(\frac{|\vec{p}|}{\kappa},\frac{\kappa}{m}))}
{2^{\la}\,\la!}\ ,
\eqe
\eq
d'=2[D'(2n+2l)+\sup(D'+1-2n-|w|,0)]\ ,
\quad
\ell'(n,l) = 2l+n-1\ .
\label{d'l'}
\eqe
\end{thm}

\noindent
{\em Proof:}

\noindent
To prove this theorem, we use the FE's for normal
products
with two insertions, given above in
eq.~\eqref{fe2ins}.
We apply a derivative $\partial^w_{\vec{p}}$ to both sides of the
equation.
Then we
integrate the FE's over $\Lambda$ subject to the appropriate
boundary condition,
using the same inductive scheme as described in the
previous subsections.
Depending on the boundary condition, we again have to distinguish the
cases
$2n+|w| \le D'$ and $2n+|w|
>D'$.  The right side of the FE, eq.~\eqref{fe2ins}, has three terms.
The first two terms involve the CAG's with two insertions, whereas the last
term only involves the CAG's with one insertion, for which we already
have
the bounds in Theorem~\ref{thm2}. The
structure of the bound for the CAG's with two insertions claimed in
the
 theorem is exactly the same
as that for one insertion, and the first two terms on the right
side of the FE also have exactly the same structure as the corresponding two
terms in the FE for one insertion. Therefore, in view of
Cor.~\ref{newcor},
the first two terms in the FE
can be treated in literally the same manner as in the previous
section with $D=D'$ there.
The third term on the right side of the FE has a different form, but
it involves only the CAG's with one insertion, for which we already have
bounds. Thus, we can concentrate only on the third term on the right
side of the FE,
and we need to show that this term satisfies our inductive bound.
We begin with the following

\begin{lemma}
For $|w| \le D'+1$, $n+1=n_1+n_2$, $l=l_1+l_2$, we have the following bound:
\bena\label{99}
&&\Bigg| \partial_{\vec p}^w \int_k {\cal
  L}^{\La,\Lao}_{2n_1,l_1}(\O_A; k, p_1,\ldots,p_{2n_1-1})\,
{\dot C}^{\La}(k)\,\,
{\cal L}^{\La,\Lao}_{2n_2,l_2}(\O_B; -k, p_{2n_1},\ldots,p_{2n})
\Bigg|\non\\
&\le&
 M\ K_1^{(4n+8l-4)|w|}\
 K_1^{D'(n+2l)^3 }\ \sqrt{|w|! \ |w'|! \ |w''|!} \ \La^{D'-2n-|w|-1} \
\e^{-m^2/\Lambda^2} \non\\
&& \times \ \sum_{\mu=0}^{d'(n,l,w,D')}\frac{1}{\sqrt{\mu!}}\
 (\frac{|\vec p|}{\La})^{\mu}\,
\sum_{\la=0}^{\la =\ell'}
\frac{\log^{\la}(\sup(\frac{|\vec{p}|}{\kappa},\frac{\kappa}{m})) \
+ \sqrt{\la !}}
{2^{\la}\,\la!}\ .
\eena
Here $K_1$ is the constant from Theorem~\ref{thm2}, and $M=5^{|w|}
2^{|w|/2} 2^{2d'}(\ell'+1)2^{\ell'+1}$.
\end{lemma}
{\it Proof:} We can pull the $\partial^w_{\vec p}$ inside the
integral. Then we first use the transformation
properties of the CAG's under translations to write
\bena
&&\partial^w_{\vec{p}}[{\cal
  L}^{\La,\Lao}_{2n_1,l_1}(\O_A(x); k, p_1,\ldots,p_{2n_1-1})\
\dot C^\Lambda(k) \
{\cal L}^{\La,\Lao}_{2n_2,l_2}(\O_B(0); -k,
p_{2n_1},\ldots,p_{2n})]
\non \\
&=& \sum_{w_1+w_2+w_3=w} c_{\{w_i\}} \
\partial^{w_3}_{\vec p} \e^{ix(k+p_1+...+p_{2n_1-1})} \
\partial^{w_1}_{\vec p}
{\cal
  L}^{\La,\Lao}_{2n_1+1,l_1}(\O_A(0); k, p_1,\ldots,p_{2n_1-1})\non\\
&&   \hspace{2cm}  \times \ \ \dot C^\La(k) \
\partial^{w_2}_{\vec p}
{\cal L}^{\La,\Lao}_{2n_2,l_2}(\O_B(0); -k, p_{2n_1},\ldots,p_{2n})
\eena
Now, the $\vec p$-derivatives on $\partial^{w_3}_{\vec p}
\e^{ix(k+p_1+...+p_{2n_1-1})}$
can be converted into $k$-derivatives, and then in the subsequent
$k$-integral
in ~\eqref{99}
moved onto the other terms by means of a partial integration, because
the integrand
decays sufficiently rapidly for large $k$ by the bounds in Theorem~\ref{thm2}.
We can then insert these bounds, and we can also use the standard multi-nomial
bound $c_{\{w_i\}} \le 3^{|w|}$.
Carrying out these manipulations, and using also \eqref{w},
and the inequality $D'(n+2l)^3 \ge [A](n_1 + 2l_1)^3 + [B](n_2 + 2l_2)^3$
in view of~\eqref{inviewof}, results in the bound:
\bena\label{101}
&&\Bigg| \partial_{\vec p}^w \int_k {\cal
  L}^{\La,\Lao}_{2n_1,l_1}(\O_A; k, p_1,\ldots,p_{2n_1-1})\,
{\dot C}^{\La}(k)\,\,
{\cal L}^{\La,\Lao}_{2n_2,l_2}(\O_B; -k, p_{2n_1},\ldots,p_{2n})
\Bigg|\non\\
&\le&
  M_0 \ K_1^{(4n+8l-4)|w|}\
 K_1^{D'(n+2l)^3 }\ \sqrt{|w|! \ |w'|! \ |w''|!} \ \La^{D'-2n-|w|-5}
\ \e^{-m^2/\Lambda^2} \non\\
&& \times \ \sum_{\mu=0}^{d_1+d_2}\frac{1}{\sqrt{\mu!}}\ \int_{k}
\e^{-|k|^2/2\Lambda^2} (\frac{|\vec p|+|k|}{\La})^{\mu}\,
\sum_{\la=0}^{\ell'}
\frac{\log^{\la}(\sup(\frac{|\vec{p}|+|k|}{\kappa},\frac{\kappa}{m}))}
{2^{\la}\,\la!}\ ,
\eena
for the constant $K_1$ provided by Theorem~\ref{thm2}, and
$M_0=5^{|w|} 2^{\ell'+1} 2^{d_1+d_2} (\ell'+1) 2^{|w|/2}$.
Here, $n+1=n_1+n_2$, $l=l_1+l_2$, and $d_1 = [A](2n_1+2l_1)
+ \sup([A]+1-2n_1-|w_1|,0)$,
$d_2 = [B](2n_2+2l_2) + \sup([B]+1-2n_2-|w_2|,0)$
is as in Theorem~\ref{thm2}. Using that $|w| \le D'+1$, we can show
that $d_1 + d_2 \le d'$, so
we can replace the upper limit of the sum over $\mu$ by $d'$.
Furthermore, we can
now bound the $k$-integral in \eqref{101} exactly as in
\eqref{sep2}
and \eqref{sep21}, leading to the
bound
\bena
&&\sum_{\mu=0}^{d'} \int_{k/\La}
\e^{-|k|^2/2\Lambda^2} (\frac{|\vec p|+|k|}{\La})^{\mu}\,
\sum_{\la=0}^{\ell'}
\frac{\log^{\la}(\sup(\frac{|\vec{p}|+|k|}{\kappa},\frac{\kappa}{m}))}
{2^{\la}\,\la!}\non\\
&\le&
2^{d'}\ \sum_{\mu=0}^{d'(n,l,w,D')}\frac{1}{\sqrt{\mu!}}\
 (\frac{|\vec p|}{\La})^{\mu}\,
\sum_{\la=0}^{\ell'}
\frac{\log^{\la}(\sup(\frac{|\vec{p}|}{\kappa},\frac{\kappa}{m}))
+ \sqrt{\la !}}
{2^{\la}\,\la!}\ .
\eena
Inserting this into the previous bound gives the statement of the lemma.
\qed

We now return to the inductive step, which consists in integrating
$\partial^w_{\vec p}$ on the
right side of the FE eq.~\eqref{fe2ins} against $\Lambda$, subject
to appropriate boundary
conditions. Concerning these boundary conditions, we must as usual
consider separately two cases:

\medskip
\noindent
{\em The case $2n+|w| > D'$:} In this case the boundary condition is
$\partial^w \L^{\Lao,\Lao}_{2n,l,D'}(\O_A
\otimes \O_B; \vec{p}) = 0$, so we integrate the right side of the FE
differentiated by
$\partial_{\vec{p}}^w$ from $\Lao$ down to $\La$. We have to consider
the three terms on the right side
separately. The first two can be handled as in the previous
subsection. So we need to focus only
on the  third term on the right side of the FE. Using the
previous lemma, and the inequality~\eqref{log}, we get
\bena
&&\Bigg| \int_\La^{\Lao} d\La' \partial_{\vec p}^w \int_k {\cal
  L}^{\La',\Lao}_{2n_1,l_1}(\O_A; k, p_1,\ldots,p_{2n_1-1})\,
{\dot C}^{\La'}(k)\,\,
{\cal L}^{\La',\Lao}_{2n_2,l_2}(\O_B; -k, p_{2n_1},\ldots,p_{2n})
\Bigg|\non\\
&\le&
 M\ K_1^{(4n+8l-4)|w|}\
 K_1^{D'(n+2l)^3 }\ \sqrt{|w|! \ |w'|! \ |w''|!} \ \int^{\Lao}_\La
 d\La' \La'^{-1+D'-2n-|w|}
\ \e^{-m^2/\Lambda'^2} \non\\
&& \times \ \sum_{\mu=0}^{d'(n,l,w,D')}\frac{1}{\sqrt{\mu!}}\
 (\frac{|\vec p|}{\La'})^{\mu}\,
\sum_{\la=0}^{\ell'}
\frac{\log^{\la}(\sup(\frac{|\vec{p}|}{\kappa'},\frac{\kappa'}{m}))
+ \sqrt{\la !}}
{2^{\la}\,\la!} \non\\
&\le& 10 M\ K_1^{(4n+8l-4)|w|}\
 K_1^{D'(n+2l)^3 }\ \sqrt{|w|! \ |w'|! \ |w''|!} \ \La^{D'-2n-|w|}  \non\\
&& \times \ \sum_{\mu=0}^{d'(n,l,w,D')}\frac{1}{\sqrt{\mu!}}\
 (\frac{|\vec p|}{\La})^{\mu}\,
\sum_{\la=0}^{\ell'}
\frac{\log^{\la}(\sup(\frac{|\vec{p}|}{\kappa},\frac{\kappa}{m}))}
{2^{\la}\,\la!} \ .
\eena
In order to bound the $\Lambda'$-integral of the third term on the
right
side of the FE,
we must additionally multiply this by $4n_1n_2$, apply the
symmetrization operator
$\mathbb S$, and sum over $n_1,n_2,l_1,l_2$, subject to
$n+1=n_1+n_2,l=l_1+l_2$.
Then we see that we reproduce the inductive bound in Theorem~\ref{thm3}
on the FE multiplied
by $1/8$ by choosing  $K \ge K_1$  sufficiently large such that
\ben
40 (l+1)(n+1)^3 \,M\, K_1^{D'(n+l)^3} \le \frac{1}{8} K^{D'(n+l)^3}\ ,
\een
which is possible in view of $D'\ge 2n+|w|+1$.

\medskip
\noindent
{\em The case $2n+|w| \le D'$:} In this case the boundary
condition is $\partial^w \L^{0,\Lao}_{2n,l,D'}(\O_A
\otimes \O_B; \vec{0}) = 0$, so we integrate the right side of the FE
differentiated by
$\partial_{\vec{p}}^w$ from $\La$ down to $0$. We have to consider the
three
terms on the right side
separately. The first two can be handled as in the previous
subsection. So we need to focus again only
on the  third term on the right side of the FE.
This is done first for
zero momentum
$\vec{p} = \vec{0}$, and the results for arbitrary momentum are then
constructed using the
Taylor formula with remainder. Using the
previous lemma, we now get
\bena\label{bound11}
&&\Bigg| \int_0^{\La} d\La' \partial_{\vec p}^w \int_k {\cal
  L}^{\La',\Lao}_{2n_1,l_1}(\O_A; k, 0,\dots,0)\,
{\dot C}^{\La'}(k)\,\,
{\cal L}^{\La',\Lao}_{2n_2,l_2}(\O_B; -k, 0,\dots,0)
\Bigg|\non\\
&\le&
 M\ K_1^{(4n+8l-4)|w|}\
 K_1^{D'(n+2l)^3 }\ \sqrt{|w|! \ |w'|! \ |w''|!}  \non\\
&& \times \ \ \int^{\La}_0 d\La' \La'^{D'-2n-|w|-1} \
\e^{-m^2/\Lambda'^2}
\sum_{\la=0}^{\ell'}
\frac{\log^{\la}(\frac{\kappa'}{m}) + \sqrt{\la !}}
{2^{\la}\,\la!} \ ,
\eena
noting that there is no boundary term. The $\La'$-integral is now bounded by
\bena
&&\int^{\La}_0 d\La' \La'^{D'-2n-|w|-1} \ \e^{-m^2/\Lambda'^2} \
\sum_{\la=0}^{\ell'} \frac{\log^\lambda(\frac{\kappa'}{m}) + \sqrt{\la !}}
{2^{\la}\,\la!}\non\\
&\le&
\La^{D'-2n-|w|} \sum_{\la=0}^{\ell'} \frac{1}{{2^{\la}\,\la!}}
\begin{cases}
\frac{1}{\la+1} \log^{\la+1}(\frac{\kappa}{m}) +
\sqrt{\la!}\log(\frac{\kappa}{m}) + 1 & \text{if $D'-2n-|w|=0$,}\\
\frac{\log^{\la}(\frac{\kappa}{m}) + \sqrt{\la !}}{D'-2n-|w|} &
\text{if $D'-2n-|w|>0$}
\end{cases} \non\\
&\le& 6(l+1) \La^{D'-2n-|w|}
\sum_{\la=0}^{\ell'+1} \frac{\log^{\la}(\frac{\kappa}{m})}
{2^{\la}\,\la!}\ .
\eena
In order to bound the $\Lambda'$ integral of the third term on the
right side of the FE,
we must additionally multiply \eqref{bound11} by $4n_1n_2$, apply
the
symmetrization operator $\mathbb S$, and sum over
$n_1,n_2,l_1,l_2$,
subject to $n+1=n_1+n_2,l=l_1+l_2$.
Then we see that we reproduce the inductive bound in
Theorem~\ref{thm3} on the FE multiplied
by $1/8$ provided that
\ben
4 \cdot 6(l+1)(\ell'+1)\ (n+1)^3 M K_1^{D'(n+l)^3}
\le \frac{1}{8} K^{D'(n+l)^3},
\quad K_1 \le K
\een
which can be satisfied for $K$ sufficiently large in view of
$|w| \le D'+1$. The bounds
at non-zero momentum are obtained using
the Taylor expansion with remainder technique as in eq.~\eqref{dev}, but now
for two insertions. The arguments are  in parallel with the case of
one insertion, noting that $d'$ satisfies the key property
$d'(D',n,l,w+v) + |w| \le d'(D',n,l,v)$ analogous to \eqref{ikey}.
Thus, each of the two terms in eq.~\eqref{dev} satisfies the inductive bound
multiplied by $1/8$ for sufficiently large $K$.

\medskip

Hence, we have seen that $\partial^w_{\vec p}$ of the third term in
the FE~\eqref{fe2ins}, integrated
against $\Lambda$, can be estimated by
$1/2$ of the inductive bound. The first two terms can be
treated in the
same manner as the corresponding terms in the FE with one insertion,
and can thereby be bounded
by $1/2$ times the inductive bound as well for sufficiently large
$K$.
This concludes the proof of the theorem.
\qed

We can insert the bound obtained in the previous theorem one
more time into the FE's
and integrate from  $0$ to $m\,$. If this is done, and if we also
carry out the sum
over $\mu$ in the bound, we obtain,
in the same way as Corollary \ref{corr1} was
obtained\footnote{The factor $\sqrt{(2n-D')_+!}\,$
can be absorbed by choosing $K$  slightly larger.}
from
Theorem \ref{thm2}:

\medskip
\begin{cor}\label{cor2}
There exists a constant $\,K>0\,$ such that:
\ben
|{\cal L}_{2n,l,D'}^{0,\Lao}(\O_A(x) \otimes \O_B(0);\vec{p})|
\le m^{D'-2n} \,
 K^{D'(n+2l)^3 }\
\sqrt{\ |w'|! \ |w''|!} \
 \sup( 1,\frac{|\vec p|}{m} )^{d'} \ \sum_{\la=0}^{2l+n}
\frac{\log^{\la}_+(\frac{|\vec{p}|}{m})}{2^{\la}\,\la!} \ .
\een
Here, $d'=2[D'(2n+2l)+\sup(D'+1-2n,0)]$.
\end{cor}

\section{Bound on the remainder in the OPE}

The bounds on the CAG's obtained in the previous section
put us in a position to give the proof of our main result, namely the bound
on the remainder of the OPE. As stated in the introduction, we
introduce test-functions $f_{p_i}$ chosen such that the
support $\supp \hat f_{p_i} \subset \{q \in \mr^4 \mid |p_i-q| \le \epsilon \}$
for some arbitrary but fixed $\epsilon$. In terms of these
test-functions, the spectator fields are defined as $\varphi(f_{p_i}) =
\int d^4 x \ \varphi(x) \ f_{p_i}(x)$.
We use the same notation as in the previous
section concerning the composite fields: $D'=[A]+[B]$,
and $A=\{n',w'\}, B=\{n'',w''\}$.
Our result, which we presented already in the introduction, is

\begin{thm}
\label{remainder}
Let the sum $\sum_C$ in the operator product expansion~\eqref{ope1} be
over all $C$ such that
\ben
[C] - [A] - [B] \le \Delta
\een
where $\Delta$ is some positive integer. Then for each such $\Delta$,
we have the following bound
for the ``remainder'' in the OPE \emph{in loop order $l$}:
\bena
&& \Bigg| \bigg\langle
\O_A(x) \O_B(0) \, \varphi(f_{p_1}) \cdots \varphi(f_{p_n}) \bigg\rangle
- \sum_{C} \C_{AB}^C(x) \,
\bigg\langle \O_C(0) \, \varphi(f_{p_1}) \cdots \varphi(f_{p_n})
\bigg\rangle \Bigg|\non\\
&& \hspace{1.5cm} \le \ \  \ m^{[A]+[B]+n}\ \sqrt{[A]![B]!}
\ \tilde K^{[A]+[B]}
\ \prod_i \sup |\hat f_{p_i}|
 \non\\
&& \hspace{1cm} \times  \
\  \sup(1,\frac{|\vec p|_n}{m})^{2([A]+[B])(n+2l+1)+3n} \
\sum_{\lambda=0}^{2l+n/2} \frac{\log^\lambda \sup(1,\frac{|\vec p|_n}{m})}
{2^\lambda \lambda!} \non\\
&& \hspace{2cm}
\times \ \ \frac{1}{\sqrt{\Delta!}} \ \Bigg( \tilde K \ m \ |x| \
\sup (1, \frac{|\vec p|_n}{m})^{n+2l+1} \Bigg)^{\Delta}\ .\non
\eena
Here, there are $n$ spectator fields, $\langle \, . \, \rangle$ denote
Schwinger functions, and $\tilde K$ is a constant depending on $n,l$.
Furthermore,
$|\vec p|_n$ is defined in eq.~\eqref{pdef}.
\end{thm}

\noindent
{\em Proof:} \\
Let us begin by defining the ``remainder functional'' for $D=0,1,2,...$ by
\bena
R^{\La,\Lao}_{D}(\O_A(x) \otimes \O_B(0))
&:=& \hbar L^{\La,\Lao}(\O_A(x) \otimes \O_B(0))-
L^{\La,\Lao}(\O_A(x))L^{\La,\Lao}(\O_B(0)) - \non \\
&& - \sum_C \C^C_{AB}(x) \ L^{\La,\Lao}(\O_C(0))  \ ,
\eena
where the sum is over all $C$ with $[C]\le D$. The corresponding moments
of this functional are written as $\mathcal{R}^{\La,\Lao}_{D,n,l}$.
Going through the definitions given in sec.~\ref{sec2}, we
can write the remainder in the OPE - with UV-cutoff $\Lao$
and IR-cutoff set to $\Lambda=0$, and still considering the full
formal power series in $\hbar$ - as
\bena\label{Rremain}
&&\bigg\langle
\O_A(x) \O_B(0) \, \hat \varphi(p_1) \cdots \hat \varphi(p_n)
\bigg\rangle - \sum_{C} \C_{AB}^C(x) \,
\bigg\langle \O_C(0) \, \hat \varphi(p_1) \cdots \hat \varphi(p_{n})
\bigg\rangle \ =\\
&-& \!\!\!\!\!\!\!\!\!\!\!\sum_{
{
\begin{array}{c}
\\[-.9cm]_{I_1 \cup ... \cup I_j = \{1,...,n\},}\\[-.3cm]
_{l,\ l_1 + ... + l_j = l}
\end{array}
}
}\!\!\!\!\!\!\!\!\! \hbar^{n+l+1-j} \
{\mathcal R}_{D,|I_1|,l_1}^{0,\Lao}(\O_A(x) \otimes \O_B(0);
\vec p_{I_1})  \ \tilde \L^{0,\Lao}_{|I_2|,l_2}(\vec p_{I_2}) \cdots
\tilde \L^{0,\Lao}_{|I_j|,l_j}(\vec p_{I_j})
\ \prod_{i=1}^n C^{0,\Lao}(p_i) \, . \non
\eena
Here, $\tilde \L^{0,\Lao}_{n,l}(\vec p)$ are the expansion coefficients of
the generating functional
$\tilde L^{0,\Lao}(\varphi) = - L^{0,\Lao}(\varphi) + \frac{1}{2}
\langle \varphi,
(C^{0,\Lao})^{-1} \star \varphi \rangle$, without any momentum
conservation delta-functions
taken out.
Thus, we need to estimate the quantities $\L^{0,\Lao}_{n,l}$,
the quantities ${\mathcal R}^{0,\Lao}_{D,n,l}$,
and the covariances $C^{0,\Lao}$. Our bounds on the CAG's without insertions
give us
\ben\label{Lestim}
| \L^{0,\Lao}_{2n,l}(\vec{p}) | \le m^{4-2n} K^{2n+4l-4} (n+l-1)!
\ \sum_{\lambda=0}^l
\frac{\log^\lambda \sup(1,\frac{|\vec{p}|}{m})}{2^\lambda
\ \lambda!}
\een
and we also have the trivial bound $C^{0,\Lao}(p)
\le[\sup(|p|,m)]^{-2}$. Thus, what remains is to give bounds on
${\mathcal R}^{0,\Lao}_{D,n,l}$.
We have the following lemma about the remainder functional:
\begin{lemma}
Let $\T^j$ be the Taylor operator introduced in
eq.~\eqref{taylor}. Then the
remainder functionals
satisfy:
\ben
R^{\La,\Lao}_{D}(\O_A(x) \otimes \O_B(0)) =(1-\Sigma_{j=0}^{\Delta} \T^j)
\left\{ \hbar L^{\La,\Lao}_D(\O_A(x) \otimes \O_B(0))-
L^{\La,\Lao}(\O_A(x))L^{\La,\Lao}(\O_B(0)) \right\} \non
\een
where $\Delta := D-D'\,$, $D'=[A]+[B]$. For $\De < 0$
 the sum is by definition empty.
\end{lemma}
{\em Proof:} Recalling the definition of the ``oversubtracted'' CAG's
with
two insertions, we
first consider the telescopic sum
\ben
L^{\La,\Lao}(\O_A \otimes \O_B) = L^{\La,\Lao}_{D}(\O_A \otimes \O_B)
+ \sum_{j=0}^{D} [L^{\La,\Lao}_{j-1}(\O_A \otimes \O_B)-
L^{\La,\Lao}_j(\O_A \otimes \O_B)] \ ,
\een
where $D<D'=[A]+[B]$. Next, for any $0 \le j$, we prove the relation
\ben\label{f1}
L^{\La,\Lao}_{j-1}(\O_A \otimes \O_B)-
L^{\La,\Lao}_j(\O_A \otimes \O_B) = \sum_{C: [C]=j} \D^C \big\{
L^{0,\Lao}_{j-1}(\O_A \otimes \O_B)
\big\} \ L^{\La,\Lao}(\O_C) \ .
\een
To see this, we make the observation that both sides of the equation
obey
the same homogeneous FE,
and the same boundary conditions, owing to the choice for the boundary
conditions
made for
the CAG's. Hence they must be equal. In view of our definition
of the OPE-coefficients
$\C_{AB}^C$ for $[C]<D'\ $ eq. (\ref{45}), we conclude that, 
for $D<D'$, we have
\ben
\hbar L^{\La,\Lao}(\O_A \otimes \O_B)
= \hbar L^{\La,\Lao}_{D}(\O_A \otimes \O_B)
+ \sum_{[C] \le D} \C^C_{AB}
\ L^{\La,\Lao}(\O_C) \ .
\een
We now subtract from both sides $L^{\La,\Lao}(\O_A)
L^{\La,\Lao}(\O_B)$,
and we bring the sum
with the OPE coefficients over to the left. Then we get the claim of
the
lemma for $D<D'$.
The case of general $D$ now works by induction. We first observe that,
for $\Delta = D-D'$, we
have
\bena\label{f2}
&&\T^{\Delta+1} \left\{ \hbar L_{D+1}^{\La,\Lao}(\O_A \otimes \O_B)
- L^{\La,\Lao}(\O_A) L^{\La,\Lao}(\O_B) \right\}\non\\
&&= -\sum_{[C]=D+1} \D^C \left\{ L^{0,\Lao}(\T^{\Delta+1} \O_A)
  L^{0,\Lao}(\O_B) \right\}
\ L^{\La,\Lao}(\O_C) \ .
\eena
This follows again because both sides satisfy the same linear,
homogeneous
FE with the same boundary
conditions. Next, using the inductive hypothesis, and making trivial
re-arrangements in the sums:
\bena
&& R_{D+1}^{\La,\Lao}(\O_A \otimes \O_B) = R_{D}^{\La,\Lao}(\O_A
\otimes \O_B)
- \Sigma_{[C]=D+1} \C^C_{AB} \ L^{\La,\Lao}(\O_C) \non\\
&=& (1-\Sigma^\Delta_{j=0} \T^j) \left\{ \hbar L_{D}^{\La,\Lao}(\O_A \otimes
  \O_B)
- L^{\La,\Lao}(\O_A) L^{\La,\Lao}(\O_B)\right\}
- \Sigma_{[C]=D+1} \C^C_{AB} \ L^{\La,\Lao}(\O_C)\non\\
&=&  (1-\Sigma^{\Delta+1}_{j=0} \T^j) \left\{ \hbar L_{D+1}^{\La,\Lao}(\O_A
  \otimes \O_B)
- L^{\La,\Lao}(\O_A) L^{\La,\Lao}(\O_B)\right\} \non\\
&+& \hbar (1-\Sigma^\Delta_{j=0} \T^j) \left\{ L_{D}^{\La,\Lao}(\O_A \otimes
  \O_B)
- L_{D+1}^{\La,\Lao}(\O_A \otimes \O_B) \right\}\non\\
&+&   \T^{\Delta+1} \left\{ \hbar L_{D+1}^{\La,\Lao}(\O_A \otimes \O_B)
- L^{\La,\Lao}(\O_A) L^{\La,\Lao}(\O_B) \right\}\non\\
&-& \Sigma_{[C]=D+1} \C^C_{AB} \ L^{\La,\Lao} (\O_C)\ .
\eena
We are now in a position to substitute the formulas~\eqref{f1}
and~\eqref{f2} for the second and third
term on the right side, together with the definition
of $\C^C_{AB}$ for $[C] \ge D'$ for the fourth term. Then the last
three terms are seen to cancel out,
and we are left with the
claim of the lemma.
\qed

Note that for a  function $f$ on $\mr^4$ of differentiability class $C^{N+1}$,
we have the formula
\ben\label{taylor1}
(1-\Sigma_{j=0}^N \T^j) f(x) = \sum_{|w|=N+1}
\frac{x^w}{N!} \int_0^1  (1-\tau)^N \ \partial^w f(\tau x)
\ d \tau \,
\een
for the remainder of a Taylor expansion in $x$ carried out to order $N$.
By~\cite{KK2}, the functionals $L_D^{\La,\Lao}(\O_A(x) \otimes \O_B(0))$
are of differentiability class $C^{\Delta}$ in the variable
$x$, where $\Delta = D-D'$, whereas the functionals
$L^{\La,\Lao}(\O_A(x))$
are smooth in $x$.
We write the operator $1-\sum_{j\le \Delta} \T^j$ in the statement of
the
previous lemma
as $(1-\sum_{j \le \Delta-1} \T^j) - \T^\Delta$, and we rewrite the first
operator in parenthesis as a remainder in a Taylor expansion to order
$N=\Delta-1$ as
in~\eqref{taylor1}. Then, by the previous lemma and the
Lowenstein-rules,
we can  write
the remainder as:
\bena\label{117}
&& \hspace{2cm} {R}^{\La,\Lao}_{D}(\O_A(x) \otimes \O_B(0)) = \sum_{|w|=\Delta}
\Bigg[\frac{x^w}{(\Delta-1)!} \int_0^1 d \tau \, (1-\tau)^{\Delta-1} \\
&& \hspace{1cm}  \ \times
\bigg( \hbar L^{\La,\Lao}_D(\partial^w \O_A(\tau x) \otimes \O_B(0))-
L^{\La,\Lao}(\partial^w \O_A(\tau x))L^{\La,\Lao}(\O_B(0))\bigg) \ -  \non\\
&& \hspace{1cm} \frac{x^w}{w!}
\bigg( \hbar L^{\La,\Lao}_D(\partial^w \O_A(0)
\otimes \O_B(0)) \ \ - \
L^{\La,\Lao}(\partial^w \O_A(0))L^{\La,\Lao}(\O_B(0))\bigg) \Bigg] \non
\eena
where $\partial^w \O_A$ on the right side denotes the linear
combination of insertions
obtained by formally carrying out the differentiations:
\ben
\partial^w \O_A = \sum_{w_1+...+w_{n'}=w} c_{\{w_i\}} \
\partial^{w_1+w'{}_1} \varphi \cdots
\partial^{w_{n'}+ w'{}_{n'}} \varphi \, .
\een
Taking the moments of this equation, and setting also $\Lambda=0$, gives:
\bena
&&\mathcal{R}^{0,\Lao}_{D,n,l}(\O_A(x) \otimes \O_B(0);
\vec{p}) = \sum_{|w|=\Delta}\Bigg[
\frac{x^w}{(\Delta-1)!} \int_0^1 d \tau \, (1-\tau)^{\Delta-1} \  \non\\
&&  \ \times \ \bigg(\mathcal{L}^{0,\Lao}_{n,l-1,D}(\partial^w \O_A(\tau x)
\otimes \O_B(0); \vec{p}) -\!\!\!\!\!\!\!\!
\sum_{
{\tiny
\begin{array}{c}
I_1 \cup I_2 = \{1,...,n\}\\
l_1 + l_2 = l
\end{array}
}
}\!\!\!\!\!\!\!\! \L^{0,\Lao}_{|I_1|,l_1} (\partial^w \O_A(\tau x);
\vec p_{I_1}) \L^{0,\Lao}_{|I_2|,l_2 } (\O_B(0);\vec p_{I_2})\bigg) \ - \non\\
&& \frac{x^w}{w!} \bigg(
\mathcal{L}^{0,\Lao}_{n,l-1,D}(\partial^w \O_A(0)
\otimes \O_B(0); \vec{p}) \ \ - \ \ \!\!\!\!\!\!\!\!\!\!
\sum_{
{\tiny
\begin{array}{c}
I_1 \cup I_2 = \{1,...,n\}\\
l_1 + l_2 = l
\end{array}
}
}\!\!\!\!\!\!\!\! \L^{0,\Lao}_{|I_1|,l_1} (\partial^w \O_A(0);
\vec p_{I_1}) \L^{0,\Lao}_{|I_2|,l_2 } (\O_B(0);\vec p_{I_2})\bigg)\Bigg]
\ . \non
\eena
At this stage, we can use our previous bounds on the CAG's to
control  the remainder. Using Cor.~\ref{cor2}, the first term in
$[\dots]$ can be bounded by
\bena
&&\Bigg| \sum_{|w|=\Delta}
\frac{x^w}{(\Delta-1)!} \int_0^1 d \tau \, (1-\tau)^{\Delta-1} \
\mathcal{L}^{0,\Lao}_{n,l-1,D}(\partial^w \O_A(\tau x) \otimes
\O_B(0); \vec{p}) \Bigg| \\
&\le& \frac{|x|^{\Delta}}{(\Delta-1)!} \sum_{|w|=\Delta}
\sup_{0 \le \tau \le 1}
\Bigg| \mathcal{L}^{0,\Lao}_{n,l-1,D}(\partial^w \O_A(\tau x)
\otimes \O_B(0); \vec{p}) \Bigg| \non\\
&& \vspace{7mm}\non\\
&\le& \frac{|x|^\Delta}{(\Delta-1)!} \ m^{D-n}
\  K^{D(n/2+2l-2)^3 }\  \sqrt{(|w'|+\Delta)! \ |w''|!} \non\\
&& \times \sum_{|w|=\Delta} \ \sum_{
w_1+...+w_{n'}=w
}
c_{\{w_i\}}
\ \sup(1,\frac{|\vec p|}{m})^{2D(n+2l-1)} \
\sum_{\lambda=0}^{2l+n/2-2} \frac{\log_+^\lambda (\frac{|\vec
    p|}{m})}{2^\lambda
\lambda!}\non\\
&\le&\ \
m^{D'-n}\  (K^{(n/2+2l-2)^3}\, m|x|)^{\Delta} \  K^{D'(n/2+2l-2)^3}
\non\\
&& \times \
 \sup(1,\frac{|\vec p|}{m})^{2D(n+2l-1)}\ \
\frac{\sqrt{\ |w'|! \ |w''|!}}{\sqrt{\Delta!}} \  \sum_{\lambda=0}^{2l+n/2-2}\
\frac{\log_+^\lambda (\frac{|\vec p|}{m})}{2^\lambda \lambda!} \ . \non
\eena
The last inequality holds for a somewhat larger constant $K$ needed
in order to absorb factors $\Delta$, 
$(4n')^{\Delta} \le (4n'+4n'')^{\Delta}\le [4(n+2l+1)]^{\Delta}$
from the sum over $w$, and 
$2^{D'+\Delta}$ from
$(|w'|+\Delta)! \le 2^{|w'|+\Delta} |w'|! \Delta! \le 2^{D'
+\Delta}|w'|!\, \Delta!\ $.
The other three terms in $[\dots]$ can be estimated in the same
 way using also our
estimates for the CAG's with one operator insertion given in
Cor.~\ref{corr1}. They 
are bounded by an expression of the same form.
Putting these straightforward estimates
together,
and defining also $\tilde K := K^{(n/2+2l)^3}$,
we thereby demonstrate the following lemma:
\begin{lemma}
Let $D' = [A]+[B], D=D'+\Delta, \Delta = 0,1,2,...$ and $A=\{n',w'\},
B=\{n'',w''\}$. The remainder functional satisfies the uniform bound
\bena
&&| \mathcal{R}^{0,\Lao}_{D,n,l}(\O_A(x) \otimes \O_B(0);
\vec{p}) | \\
&\le&m^{D'-n} \ (\tilde K \ m \ |x|)^{\Delta} \   \tilde K^{D'}
\ \sup(1,\frac{|\vec p|}{m})^{2D(n+2l+1)} \
\frac{\sqrt{\ |w'|! \ |w''|!}}{\sqrt{\Delta!}} \ \sum_{\lambda=0}^{2l+n/2}
\frac{\log_+^\lambda (\frac{|\vec p|}{m})}{2^\lambda \lambda!}\non
\eena
with a constant $\tilde K$ depending only on $n,l$.
\end{lemma}
Substituting the bound stated in the lemma into eq.~\eqref{Rremain}, using the
trivial estimate $C^{0,\Lao}(p) \le [\sup(m,|p|)]^{-2}$, the
estimate~\eqref{Lestim},
and the fact that
$\hat f_{p_i}$ is supported in a ball of radius $\epsilon$ around
$p_i\,,\ $
we get the statement
of the theorem for a sufficiently large new constant $\tilde K$.
This completes the proof.
\qed

\vspace{2cm}

\noindent
{\bf Acknowledgements:} S.H. would like to thank CPHT, Ecole
Polytechnique, for hospitality
and financial support. He also acknowledges support under
 ERC~grant~QC\&C259562.

\end{document}